\documentclass[]{article}
\usepackage[square,sort&compress,comma]{natbib}
\usepackage{graphicx}
\usepackage{multirow}
\usepackage[colorinlistoftodos]{todonotes}
\usepackage{soul}
\usepackage{rotating}
\usepackage{amssymb}
\usepackage{amsmath}
\usepackage{amsfonts}
\usepackage{mathrsfs}
\usepackage[hidelinks]{hyperref}
\usepackage{fontenc}
\usepackage{inputenc}
\DeclareMathOperator*{\argmax}{argmax}

\newcommand{\R}{\textsf{R} }
\newcommand{\loglikP}{l\left(\mathbf{P}\right)}

\newcommand{\fulltitle}{Maximum likelihood estimation for nonembeddable Markov chains when the cycle length is shorter than the data observation interval}
\author{Duncan Ermini Leaf\\~\\University of Southern California\\Leonard D. Schaeffer Center for Health Policy \& Economics\\~\\dleaf@usc.edu}

\bibpunct{[}{]}{,}{n}{ }{ }
\setcitestyle{notesep={ }}

\title{\fulltitle}

\begin{document}

\maketitle

\begin{abstract}
Time-homogeneous Markov chains are often used as disease progression models in studies of cost-effectiveness and optimal decision-making. Maximum likelihood estimation of these models can be challenging when data are collected at a time interval longer than the model's transition cycle length.  For example, it may be necessary to estimate a monthly transition model from data collected annually.  The likelihood for a time-homogeneous Markov chain with transition matrix $\mathbf{P}$ and data observed at intervals of $T$ cycles is a function of $\mathbf{P}^T.$  The  maximum likelihood estimate of $\mathbf{P}^T$ is easily obtained from the data.  The $T$th root of this estimate would then be a maximum likelihood estimate for $\mathbf{P}.$   However, the $T$th root of $\mathbf{P}^T$ is not necessarily a valid transition matrix. 
Maximum likelihood estimation of $\mathbf{P}$ is a constrained optimization problem when a valid root is unavailable.
The optimization problem is not convex.  Local convergence is explored in several case studies through graphical representations of a grid search.  The example cases use disease progression data from the literature as well as synthetic data.
The global maximum likelihood estimate is increasingly difficult to locate as the number of cycles or the number of states increases.
What seems like a straightforward estimation problem can be challenging even for relatively simple models.  Researchers should consider alternatives to likelihood maximization or alternative models.


\end{abstract}

\section{Introduction}
\label{sec:intro}

Consider a time-homogeneous Markov chain over $s$ discrete states.  These states could be disease severity stages, for example.  The transition probabilities for a single cycle are given by the $s\times s$ matrix $\mathbf{P}$, where $p_{ij} \in [0,1]$ is the probability of residing in state $j$ at the end of the cycle conditional on starting in state $i$ at the beginning of the cycle.  Now suppose that data are collected at intervals of $T$ cycles.  
If, for example, data are collected annually and each cycle is one month, then $T=12.$
The transition probabilities over the data observation interval are $\mathbf{P}^T$.  
Let $\mathbf{N}$ be the $s \times s$ matrix of observed data, where $n_{ij}$ counts the number of transitions starting in state $i$ and residing in state $j$ at the end of the observation interval.  
The maximum likelihood estimate (MLE) of $\mathbf{P}^T$ is $\widehat{\mathbf{P}^T}$ with $(i,j)$th element $n_{ij} / \sum_{k=1}^{s} n_{ik}$.\footnote{Estimates are denoted with a hat over the estimand so that $\widehat{\mathbf{P}^T}$ is an estimate of $\mathbf{P}^T,$ but $\widehat{\mathbf{P}}^T$ is the $T$th power of an estimate of $\mathbf{P}.$}  If $\widehat{\mathbf{P}^T}$ has a $T$th root that is a stochastic matrix,\footnote{A stochastic matrix has non-negative elements and rows summing to unity.} then $\widehat{\mathbf{P}} = (\widehat{\mathbf{P}^T})^{1/T}$ is an MLE of $\mathbf{P}$.
Unfortunately, $\widehat{\mathbf{P}^T}$ may not have a stochastic $T$th root.
Two examples are given below and simulation results show this occurs frequently \cite{jahn_alternative_2019}.  This report focuses on likelihood maximization when there is no stochastic root.

Chhatwal et al. \cite{chhatwal_changing_2016} present an example in which there is no stochastic root using an HIV progression model from Chancellor et al. \cite{chancellor_modelling_1997}. In this model, patients can progress through three disease states. The fourth state, death, is censoring and absorbing. 
Consider the following annual transition counts between disease states and death:\footnote{In fact, Chancellor et al. estimate annual transition probabilities using survival analysis of hospital patient data \cite{chancellor_modelling_1997}. Details on follow-up times are not given.
The counts in (\ref{eqn:chancellordata}) are estimates of expected counts assuming annual follow-ups.
See Section \ref{sec:discussion} for further discussion of survival analysis.}  
\begin{equation}
\mathbf{N} = \left[ \begin{array}{rrrr} 4494 &  1257 &  417 & 61\\
                                            0 & 1734 & 1214 & 36\\
                                            0 &   0 &  6724 & 2240\\
                                            0 &   0 &    0 & 0\end{array} \right].
\label{eqn:chancellordata}
\end{equation}
The fourth row of $\mathbf{N}$ has no transitions from death due to censoring.
The stochastic root problem arises when trying to estimate monthly transition probabilities from these annual counts.
With the constraint that death is absorbing, the MLE for annual transition probabilities is:
\[
\widehat{\mathbf{P}^{12}} = \left[ \begin{array}{rrrr} 
.721 & .202 &  .067 & .010 \\
0 &     .581 & .407  & .012\\
0 &   0 &    .750 &   .250\\
0 &   0 &    0 &   1\end{array} \right].
\]
Since the annual MLE has distinct, positive eigenvalues ($\lambda_2 = .750,$ $\lambda_3 = .721,$ and $\lambda_4=.581$),\footnote{$\lambda_1 = 1$ is omitted in any discussion of eigenvalues.} its principal 12th root can be obtained via diagonalization of the eigendecomposition,
 $(\widehat{\mathbf{P}^{12}})^{1/12} = \mathbf{A} \mathbf{D}^{1/12} \mathbf{A}^{-1}$, where $\mathbf{A}$ is the matrix of eigenvectors for $\widehat{\mathbf{P}^{12}}$
and $\mathbf{D}$ is the diagonal matrix of eigenvalues.
Unfortunately, the result is not a stochastic matrix due a negative probability:
\[
\widehat{\mathbf{P}^{12}}^{1/12} = \left[ \begin{array}{rrrr} 
.973 & .025 &  .001 & .001 \\
0 &     .956 & .049  & -.005\\
0 &   0 &    .976 &   .024\\
0 &   0 &    0 &   1\end{array} \right].
\]
Section \ref{sec:study2} provides a method for finding the stochastic MLE in this example.

The previous example has a real-valued root that is not a stochastic matrix.  Sometimes, a real-valued root is not available at all.  Consider the five-state HIV model of Sendi et al. \cite{sendi_estimating_1999} presented in Charitos et al. \cite{charitos_computing_2008}.  The final state is absorbing.  The six-month transition counts are \cite{charitos_computing_2008}:
\begin{equation}
\mathbf{N} = \left[ \begin{array}{rrrrr} 339 &  31 &  24 & 17 & 5\\
233 & 73 & 55 & 49 & 6\\
150 &  77 &  63 & 91 & 34\\
70 & 26 & 60 & 193 & 66\\
0 & 0 & 0 & 0 & 415\end{array} \right].
\label{eqn:sendicharitosdata}
\end{equation}
The MLE for six-month transition probabilities is:
\[
\widehat{\mathbf{P}^{6}} = \left[ \begin{array}{rrrrr} 
.815 & .075 & .058 & .041 & .012 \\
.560 & .175 & .132 & .118 & .014\\
.361 & .186 & .152 & .219 & .082\\
.169 & .063 & .145 & .465 & .159\\
0 &   0 &    0 &   0  &    1\end{array} \right].
\]
Here, the smallest eigenvalue is negative ($\lambda_5=-.005$) and so the sixth root of $\widehat{\mathbf{P}^{6}}$ will have an imaginary part. Section \ref{sec:study3} finds the monthly MLE for this example.

There are several existing estimation methods that work around the unavailable stochastic root.  These methods can be classified first on whether or not they assume a continuous Markov process underlies the observed data. This is known as embeddability. In the embeddable model, there is a matrix $\mathbf{G}$ such that $\mathbf{P}^T = e^{\mathbf{G}}.$
Once $\mathbf{G}$ is estimated from observed data, a stochastic $T$th root exists for any integer choice of $T$ \cite{kingman_imbedding_1962}.  Along with the embeddability question, estimation methods can be classified as likelihood maximization, Bayesian, or matrix approximations.
Table \ref{tab:methods_classification} classifies methods by embeddability and how the likelihood is used.  Lin \cite{lin_roots_2011} discusses the various methods in more detail.

The approximation methods replace invalid matrix elements or find a valid matrix that minimizes a distance to the invalid estimate.  
They do not use the likelihood beyond the initial MLE of transition probabilities over the data observation interval. 
This approach has an advantage over likelihood maximization and Bayesian methods in that it provides an estimate when transition counts are not readily available (see, e.g., \cite{jahn_alternative_2019} and \cite{israel_finding_2001}) or when   
estimated transition probabilities for multiple cycles come from some other model type, such as a survival model (e.g., \cite{chhatwal_changing_2016}).
However, unlike likelihood maximization and Bayesian methods, approximation estimates do not have a theoretical basis for statistical inference.
Section \ref{sec:discussion} discusses alternatives to the approximation approach via improved data accessibility and direct use of other model types.

This report focuses on likelihood maximization without assuming embeddability.  The assumption here  is that a stochastic $T$th root exists for the specific value of $T$ in question, but not necessarily for other choices of $T.$  This is the least restrictive form of the model.  Compared to approximation methods, likelihood maximization is desirable because it can provide an approximate sampling distribution of the transition matrix estimate.  The sampling distribution can then be used for computing confidence regions, hypothesis tests, parametric bootstrapping, and probabilistic sensitivity analyses.\footnote{See, e.g., the ISPOR-SMDM Modeling Good Research Practices Task Force recommendation on using an estimate's sampling distribution in probabilistic sensitivity analyses \cite{briggs_model_2012}.}

As shown in Table \ref{tab:methods_classification}, Craig and Sendi's \cite{craig_estimation_2002} is the only existing approach to likelihood maximization without assuming embeddability. They suggest an iterative method based on the expectation-maximization (EM) algorithm and 
recommend using several initial transition matrices because it may converge to local maxima.
However, Craig and Sendi 
do not apply their suggested method to this particular problem.  It is not clear how the local maxima arise in practice nor how best to identify a global maximum.  Furthermore, Craig and Sendi's 
approach requires effort to implement and test their EM algorithm. It is not an off-the-shelf solution.

Craig and Sendi \cite{craig_estimation_2002} apply their method to the related problem where data are observed at intervals of varying length, including an interval equal to the cycle length.  In that model, the likelihood is necessarily parameterized in terms of $\mathbf{P}$ instead of $\mathbf{P}^T$, thereby avoiding the stochastic root problem.
MacDonald \cite{macdonald2014} demonstrates  direct optimization as a simpler alternative to EM in Craig and Sendi's application. Like Craig and Sendi, MacDonald does not investigate local maxima.

This report is a case study of the multiple local maxima issue mentioned in Craig and Sendi \cite{craig_estimation_2002}.
However, this report departs from Craig and Sendi
in the direction of MacDonald \cite{macdonald2014} by using a widely-available optimization procedure instead of a bespoke EM implementation.  
Section \ref{sec:methods} lays out the general details of the optimization procedure and grid search approach to finding local convergence points.  These methods are applied to several examples with results reported in Section \ref{sec:studies}.  Section \ref{sec:discussion} has general comments about the approach. \\~\\

\begin{table}[h]
	\centering
	\begin{tabular}{|l|l|l|}
		\cline{2-3}
		\multicolumn{1}{c}{} & \multicolumn{2}{|c|}{\textbf{Embeddability assumed?}}	\\	
		\cline{2-3}
		\multicolumn{1}{c}{} & \multicolumn{1}{|c|}{\textbf{No}} & \multicolumn{1}{|c|}{\textbf{Yes}}	\\	
		\hline
		\multirow{2}{1.05in}{\textbf{Likelihood maximization}} & \multirow{2}{1.05in}{Craig and Sendi \cite{craig_estimation_2002}, \textbf{this report}} &  
		\multirow{2}{2.15in}{Bladt and S{\o}rensen \cite{bladt_statistical_2005}, Kalbfleisch and Lawless \cite{kalbfleisch_analysis_1985}} \\
		& & \\
		\hline
		\multirow{2}{1.05in}{\textbf{Bayesian}} & & \multirow{2}{2.15in}{Bladt and S{\o}rensen \cite{bladt_statistical_2005}, Welton and Ades \cite{welton_estimation_2005}} \\
		& & \\
		\hline
		\multirow{5}{1.05in}{\textbf{Approximation}} & \multirow{5}{1.05in}{Charitos et al. \cite{charitos_computing_2008}, Kreinin and Sidelnikova \cite{kreinin_regularization_2001}} & \multirow{5}{2.15in}{Coleman  \citetext{\citealp[chapter~5]{coleman_introduction_1964}}, 
			Crommelin and Vanden-Eijnden \cite{crommelin_fitting_2006}, 
			Israel et al. \cite{israel_finding_2001}, 
			Jahn et al. \cite{jahn_alternative_2019},
			Kreinin and Sidelnikova \cite{kreinin_regularization_2001},
			Singer and Spilerman \cite{singer_representation_1976},
			Zahl \cite{zahl_markov_1955}
		}\\
		& & \\
		& & \\
		& & \\
		& & \\
		\hline
	\end{tabular}
	\caption{Estimation methods classified on (1) whether or not they make the embeddability assumption of an underlying continuous-time Markov process and (2) how the likelihood is used.}
	\label{tab:methods_classification}
\end{table}

\section{Methods}
\label{sec:methods}

The likelihood comes from the multinomial distribution for each row of the observed data matrix, $\mathbf{N}.$  For stochastic matrix  $\mathbf{P},$ the log-likelihood is:
\begin{equation}
l\left(\mathbf{P}\right) = \sum\limits_{i=1}^{s} \sum\limits_{j=1}^{s-1} n_{ij} \log \left(\mathbf{P}^T\right)_{ij} + \sum\limits_{i=1}^{s} n_{is} \log \left(1-\sum\limits_{j=1}^{s-1} \left(\mathbf{P}^T\right)_{ij}\right).
\label{eqn:loglik}
\end{equation}
Note that the likelihood is parameterized in terms of the root, $\mathbf{P}$, instead of treating $\mathbf{P}^T$ as the parameter.  This avoids the problem of not being able to find a root.  However, this also means that, as a function of $\mathbf{P}^T$, the domain of the likelihood is restricted to $\mathcal{P}_s(T)$, the set of $s\times s$ stochastic matrices that have a stochastic $T$th root. 
For $s>2$, $\mathcal{P}_s(T)$ is not convex \cite{lin_roots_2011}.
As a function of \emph{unrestricted} $\mathbf{P}^T$, the likelihood is concave  with the unique global maximizer $\widehat{\mathbf{P}^{T}}$ discussed in Section \ref{sec:intro}. However, if $\widehat{\mathbf{P}^{T}} \not\in \mathcal{P}_s(T)$, then the maximizer must be found elsewhere and there 
could potentially be multiple local maxima due to the non-convexity of $\mathcal{P}_s(T)$.

Since the rows of $\mathbf{P}$ must sum to unity, the problem is further 
parameterized in terms of the $s \times s-1$ matrix $\boldsymbol{\Theta}$ in order to remove the redundant final column:
\[
\mathbf{P} = \left[\begin{array}{ccc} \boldsymbol{\Theta} && \mathbf{1}_s - \boldsymbol{\Theta}\mathbf{1}_{s-1} \end{array}\right],
\]
where $\mathbf{1}_s$ is the $s$-dimensional column vector of ones.
Each row of $\boldsymbol{\Theta}$ is a point in the $(s-1)$-simplex.
The goal then is to maximize $l\left(\mathbf{P}\right)$ with respect to $\boldsymbol{\Theta}$.

Each case study of Section \ref{sec:studies} follows the same sequence of steps. The first step confirms that the observation interval MLE does not have a stochastic root (except for Study 1, which has a known root).  Section \ref{sec:rootfinding} gives the general approach to root-finding.  Next, a grid search is used to locate the local convergence points of an optimization procedure.  The first step in the grid search generates a large number, $M$, of equally spaced values for $\boldsymbol{\Theta}$.
This is described in Section \ref{sec:gridsearch}.  Each of these matrices is used as the starting point for the iterative optimization described in Section \ref{sec:optim}.  Each optimization runs until its convergence criteria are achieved. This produces a set of convergence points $\widetilde{\mathbf{P}}_1, \ldots, \widetilde{\mathbf{P}}_M$.  Lastly, the convergence points are analyzed in order to identify a global maximizer of the likelihood.  The analysis strategy for convergence points is described briefly in Section \ref{sec:convanalysis} and demonstrated through examples in Section \ref{sec:studies}.

\subsection{Root-finding}
\label{sec:rootfinding}

In each Section \ref{sec:studies} study, the MLE over the observation interval is nonsingular and has distinct eigenvalues, so only primary stochastic roots are of interest. In Study 1, the stochastic root is known already.  The first question in the remaining studies is whether there are any \emph{real-valued} (not necessarily stochastic) primary roots.  Theorem 2.4 of \cite{higham_pth_2011} provides the answer.  Each study considers only even values of $T.$  
With $T$ even, there are no real-valued roots at all when a negative eigenvalue is present (Studies 3, 4, 5, 7, and 8).  
When there are no negative eigenvalues (Studies 2 and 6), let $r$ be the number of positive eigenvalues and $c$ be the number of complex conjugate eigenvalue pairs.  The number of real-valued primary roots is $2^rT^c$, some of which might be stochastic matrices.  Higham and Lin discuss  necessary conditions for the existence of a stochastic root \cite{higham_pth_2011}.  In Studies 2 and 6, the real-valued roots are enumerated and inspected for stochasticity.

\subsection{Grid setup}
\label{sec:gridsearch}

Consider the case when there are no additional parameter constraints beyond  $\mathbf{P}$ being a stochastic matrix.  A grid is created for a single row of  $\boldsymbol{\Theta}$ by taking a set of equally spaced points in the interior of the $(s-1)$-dimensional unit hypercube and intersecting them with the $(s-1)$-simplex. 
The grid for the $\boldsymbol{\Theta}$ matrix is formed from the Cartesian product of $s$ copies of the single-row grid point set.

Some of the studies in Section \ref{sec:studies} have fixed parameter values determined by absorbing states or by the disease progression model.  For these cases, each single-row grid point set is formed by creating a set of equally spaced points for the free parameters in the row, fixing the constrained parameters in the row, and intersecting with the $(s-1)$-simplex.  The grid for $\boldsymbol{\Theta}$ is then formed from the Cartesian product of these single-row sets.

\subsection{Optimization method}
\label{sec:optim}

Each $\boldsymbol{\Theta}$ matrix in the grid is used as a starting point for iterative optimization of the log-likelihood  subject to the constraints: $\theta_{ij} \geq 0$ ($i=1,\ldots,s$, $j=1,\ldots,s-1$) and $\sum_{j=1}^{s-1} \theta_{ij} \leq 1$ for $i=1,\ldots,s$.  The \emph{constrOptim} function in \R (version 3.6.3)enforces the constraints by adding  a $\log$ barrier function to the log-likelihood \citetext{\citealp[chapter~16]{lange2010}, \citealp{Rlang}}.  Outer iterations reduce the barrier function toward zero at each step so that the objective approaches the log-likelihood.

Within each outer iteration, BFGS optimization seeks a local maximizer of the combined log-likelihood and barrier function.  The barrier gradient is described in \cite[chapter~11]{boyd_vandenberghe_2004}.  When there are no fixed parameter values, the gradient of the log-likelihood (\ref{eqn:loglik}) is a matrix $\nabla_\theta \loglikP$ with $(u,v)$th element:
\begin{equation}
\frac{\partial \loglikP}{\partial \theta_{uv}} = \sum\limits_{i=1}^{s} \sum\limits_{j=1}^{s-1} \left[ \frac{n_{ij}}{\left(\mathbf{P}^T\right)_{ij}} -  \frac{n_{is}}{\left(1-\sum\limits_{k=1}^{s-1} \left(\mathbf{P}^T\right)_{ik}\right)} \right] \frac{\partial \left(\mathbf{P}^T\right)_{ij}}{\partial \theta_{uv}}.
\label{eqn:loglikgradient}
\end{equation}
Each $\partial \left(\mathbf{P}^T\right)_{ij} / \partial \theta_{uv}$ term in (\ref{eqn:loglikgradient}) can be obtained computationally as,
\begin{equation}
\frac{\partial \mathbf{P}^T}{\partial \theta_{uv}} = \sum\limits_{k=1}^{T} \mathbf{P}^{k-1} \frac{\partial \mathbf{P}}{\partial \theta_{uv}} \mathbf{P}^{T-k},
\label{eqn:delPT_delp}
\end{equation}
for $u=1,\ldots,s$ and $v=1,\ldots,s-1$, where 
\[
\left(\frac{\partial \mathbf{P}}{\partial \theta_{uv}}\right)_{ij} = \left\{ \begin{array}{rcl}  1 && \mbox{if $i=u$ and $j=v$;}\\
-1 && \mbox{if $i=u$ and $j=s$;}\\
0 && \mbox{otherwise,}	  \end{array}\right.
\]
and $\mathbf{P}^0$ is the identity matrix.  Thus,
the matrix in (\ref{eqn:delPT_delp}) has $(i,j)$th element,
\[
\frac{\partial \left(\mathbf{P}^T\right)_{ij}}{\partial \theta_{uv}} = \sum\limits_{k=1}^T \left(\mathbf{P}^{k-1}\right)_{ui} \left[\left(\mathbf{P}^{T-k}\right)_{jv} - \left(\mathbf{P}^{T-k}\right)_{sv}\right].
\]
Gradient entries can be set to zero for $\theta_{uv}$ parameters that are fixed constants determined by the model structure.

BFGS optimization stops when either the absolute change or the relative change in the objective is below some given tolerance.  The outer iteration is considered to have converged once the relative change between iterations is below a given tolerance.

\subsection{Analysis of convergence points}
\label{sec:convanalysis}

When the optimization at each grid location $i$ reaches its convergence criteria, the output is a stochastic matrix $\widetilde{\mathbf{P}}_i$. The next step is to find the global maximizer $\widehat{\mathbf{P}}$ among these convergence points:
\[
\widehat{\mathbf{P}} = \argmax_{i}\, l\left(\widetilde{\mathbf{P}}_i\right).
\]  
This can give multiple values of $\widehat{\mathbf{P}}$.  There are two possible reasons for this. First, a matrix could have multiple $T$th roots, i.e., $\widetilde{\mathbf{P}}_i \neq \widetilde{\mathbf{P}}_j$, but $\widetilde{\mathbf{P}}_i^T = \widetilde{\mathbf{P}}_j^T$ for some $i \neq j$.  Second, there could be multiple distinct maximizers, i.e.,  $l\left(\widetilde{\mathbf{P}}_i\right) = l\left(\widetilde{\mathbf{P}}_j\right)$, but $\widetilde{\mathbf{P}}_i^T \neq \widetilde{\mathbf{P}}_j^T$ for some $i \neq j$.  In order to distinguish between these two cases, the search looks for the global maximizer of the likelihood as a function of $\widetilde{\mathbf{P}}_i^T$ for each $i=1,\ldots,M$. 
If there is only one global maximizer, then each of its roots is a valid MLE.  If there are multiple global maximizers, then each of their roots are also all valid MLEs.
Additionally, looking for the global maximizer in terms of $\widetilde{\mathbf{P}}_i^T$, and then calculating all of the stochastic $T$th roots of $\widetilde{\mathbf{P}}_i^T$ allows for the possibility that the grid search does not find one of the roots.  This becomes more important as the dimensionality of the problem grows and fewer grid points are available.

\section{Numerical studies}
\label{sec:studies}

This section presents eight numerical studies that demonstrate the methods from Section \ref{sec:methods}.  The first three use data from published studies. The first has a known solution while the second and third are unsolved problems. The remaining studies are synthetic examples designed to explore performance in varying dimensions ($s=3$ and $4$), eigenstructures, and observation interval lengths ($T=2,$ $24,$ and $100$).

Table \ref{tab:study_summary} shows the number of free parameters and grid sizes for each study.  Each grid is equally divided between multiple processors and the optimizations are run in parallel. Study 3 uses 64 processors while all other studies use 20 processors.  
The grid sizes are determined by the number of states ($s$), limits of the \emph{data.table} \cite{data.table} structure in \R, number of parallel processes, and computing resource limits.
For example, Study 7 uses more grid points than Study 6 because Study 7 has six additional parameters. Using a finer grid in Study 6 would scale the number of grid points beyond the number of grid points in Study 7, but also beyond computing resource limitations.

Table \ref{tab:study_summary} also shows the convergence tolerances for each study.  The global maximizer is at the boundary of the parameter space in every study except the first. In other words, there are zeroes in the transition matrices.  
Smaller convergence tolerances cause the parameters to move further toward the boundary before BFGS optimization stops.
As the parameters move closer to the boundary, the optimization becomes increasingly sensitive to the barrier part of the objective.    
Optimizations fail due to infinite values of the objective for some grid locations in some studies when the convergence tolerances are very small.   Therefore, convergence tolerances are chosen to be as small as possible while still allowing optimizations for most grid points to run to completion. Table \ref{tab:study_summary} shows completion rates.
A tuning parameter controls the shape of the barrier over outer iterations.  The default ($10^{-4}$) is used in all studies.
An alternative approach would use a common set of convergence tolerances across all studies and adjust the barrier tuning parameter specifically for each study.

\begin{sidewaystable}
	
	\centering
	
	\begin{tabular}{|ccccccccc|}
		\hline
		&&& \textbf{Grid} && \textbf{Outer} & \textbf{Inner}  & \textbf{Inner} & \textbf{Grid}\\		
		&&& \textbf{size}  & \textbf{Cycles} & \textbf{relative} & \textbf{absolute}  & \textbf{relative} & \textbf{completion}\\		
		\textbf{Study}  &  \textbf{Motivation} &  \textbf{Parameters} & ($M$) & ($T$) &  \textbf{tolerance} & \textbf{tolerance}  & \textbf{tolerance} & \textbf{rate} (\%)\\
		\hline
		1 & known root & 6 & 6,859,000 & 6 & $10^{-11}$ & $10^{-9}$ & $10^{-9}$ & $>99.999$  \\
		\hline
		2 & negative probability & 6 & 4,332,000 & 12 & $10^{-13}$ & $10^{-12}$ & $10^{-12}$ & 94.050 \\
		\hline
		3 & negative eigenvalue & 16 & 24,010,000 & 6 &  $10^{-12}$ & $10^{-10}$ & $10^{-10}$ & 99.892 \\
		\hline
		\multirow{3}{*}{4} & \multirow{3}{*}{negative eigenvalue} & \multirow{3}{*}{6} & \multirow{3}{*}{6,859,000} & 2 & $10^{-10}$ & $10^{-8}$ & $10^{-8}$ & $>99.999$ \\
		\cline{5-9}
		&& & & 24 & $10^{-11}$ & $10^{-9}$ & $10^{-9}$ & $>99.999$ \\
		\cline{5-9}
		&& & & 100 & $10^{-12}$ & $10^{-10}$ & $10^{-10}$ & 100 \\
		\hline
		\multirow{3}{*}{5} & \multirow{3}{*}{negative eigenvalues} & 	\multirow{3}{*}{6} & 	\multirow{3}{*}{6,859,000} & 2 & $10^{-10}$ & $10^{-8}$ & $10^{-8}$ & $>99.999$ \\
		\cline{5-9}
		&&&& 24 & $10^{-11}$ & $10^{-9}$ & $10^{-9}$ & $>99.999$ \\
		\cline{5-9}
		&&&& 100 & $10^{-12}$ & $10^{-10}$ & $10^{-10}$ & $>99.999$ \\
		\hline
		\multirow{3}{*}{6} & \multirow{3}{*}{complex eigenvalues} & \multirow{3}{*}{6} & \multirow{3}{*}{6,859,000} & 2 & $10^{-10}$ & $10^{-8}$ & $10^{-8}$ & $>99.999$ \\
		\cline{5-9}
		&&&& 24 & $10^{-11}$ & $10^{-9}$ & $10^{-9}$ & $>99.999$ \\
		\cline{5-9}
		&&&& 100 & $10^{-12}$ & $10^{-10}$ & $10^{-10}$ & $>99.999$ \\
		\hline
		\multirow{3}{*}{7} & \multirow{3}{*}{negative eigenvalue} & \multirow{3}{*}{12} & \multirow{3}{*}{9,834,496} & 2 & $10^{-10}$ & $10^{-8}$ & $10^{-8}$ & $>99.999$ \\
		\cline{5-9}
		&&&& 24 & $10^{-11}$ & $10^{-9}$ & $10^{-9}$ & $>99.999$ \\
		\cline{5-9}
		&&&& 100 & $10^{-12}$ & $10^{-10}$ & $10^{-10}$ & $>99.999$ \\
		\hline
		\multirow{3}{*}{8} & negative & \multirow{3}{*}{12} & \multirow{3}{*}{9,834,496} & 2 & $10^{-10}$ & $10^{-8}$ & $10^{-8}$ & $>99.999$ \\
		\cline{5-9}
		&and complex &&& 24 & $10^{-11}$ & $10^{-9}$ & $10^{-9}$ & $>99.999$ \\
		\cline{5-9}
		& eigenvalues&&& 100 & $10^{-12}$ & $10^{-10}$ & $10^{-10}$ & $>99.999$ \\
		\hline
	\end{tabular}
	\caption{Summary of each numerical study in Section \ref{sec:studies}. \textbf{Motivation} describes why the problem is chosen for study. 
	\textbf{Parameters} is the number of free parameters in the optimization. \textbf{Grid size} is the number of grid points at which the optimization is performed for the grid search.  \textbf{Cycles} is the number of transition cycles in the observation interval.  Tolerances are described in Section \ref{sec:studies}.  \textbf{Grid completion rate} gives the percentage of grid points in the \textbf{Grid size} column for which the optimizations reach the convergence criteria. }
	\label{tab:study_summary}
\end{sidewaystable}

\subsection{Study 1: 3-state HIV model with known stochastic root}

The first numerical study uses six-month HIV progression data from Table 1 of \cite{craig_estimation_2002}.  The goal is a monthly transition matrix. The six-month MLE, $\widehat{\textbf{P}^6}$, has a stochastic sixth root that is easily obtained via eigendecomposition.  Eigendecomposition is superior for this case because of its computational ease.  However, the known result provides a validation target for the present approach.  

\begin{figure}
\centering
\includegraphics[width=1.0\linewidth]{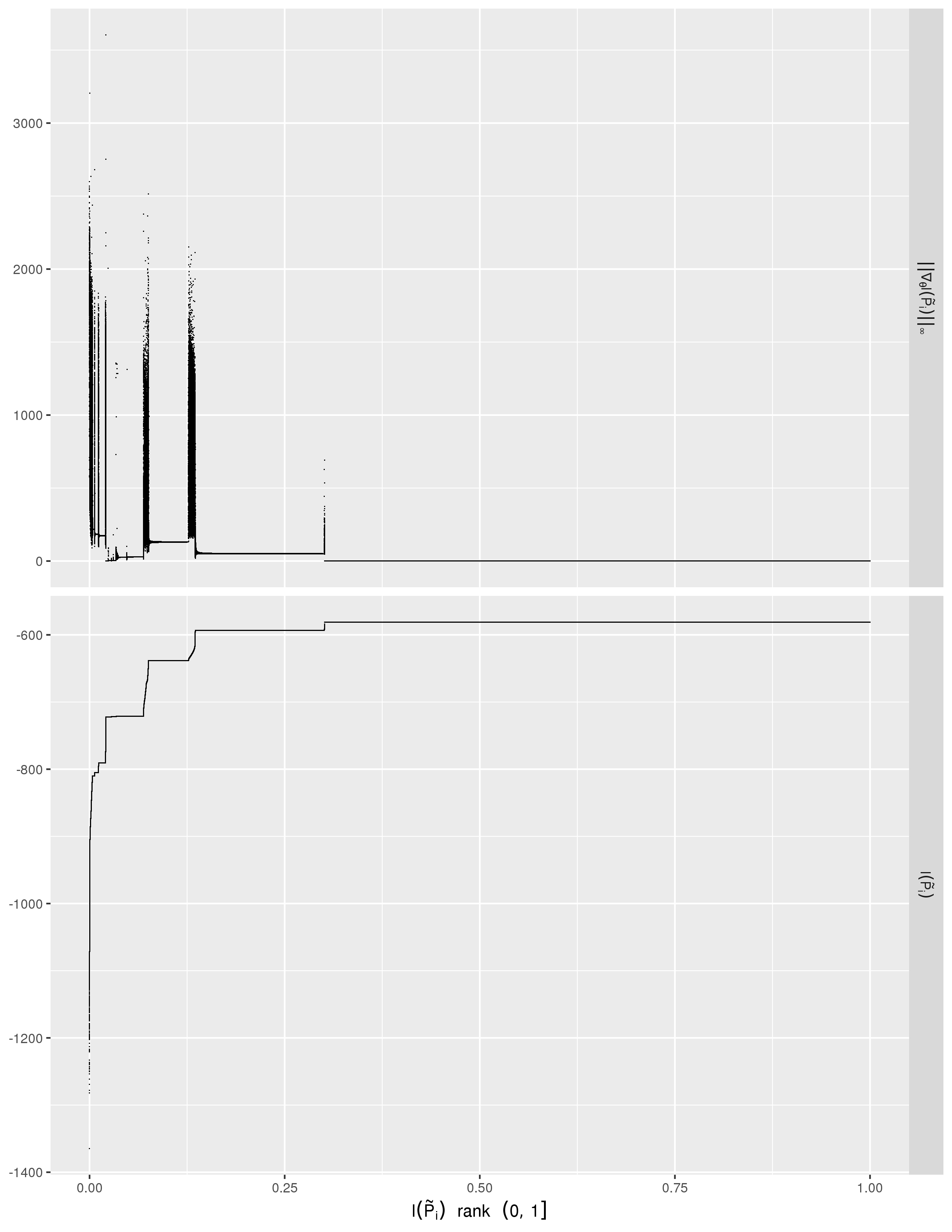}
\caption{Bottom panel: Study 1 log-likelihood evaluated at each convergence point. Top panel: $L^\infty$ norm of gradient at each convergence point. Both panels are plotted along rank order of the log-likelihood for each convergence point (horizontal axis, $(0,1]$ scale). Any ties are ranked arbitrarily.}
\label{fig:study1_sortedlik}
\end{figure}

In the bottom panel of Figure \ref{fig:study1_sortedlik}, the log-likelihood is evaluated at each convergence point, $\widetilde{\textbf{P}}_i$ ($i=1, \ldots, 6859000$) and plotted against its rank scaled to the $(0,1]$ interval.  
The optimizations for 93 grid points do not converge due to the numerical issue described previously.  These points are not represented in Figure \ref{fig:study1_sortedlik}.
The rightmost point represents the global maximum found in the grid search. The corresponding maximizer, $\widehat{\mathbf{P}},$ matches the sixth root reported in \cite{craig_estimation_2002} to four decimal places. The top panel of Figure \ref{fig:study1_sortedlik} shows the $L^\infty$ norm of the gradient.  As expected, the gradient is approximately zero near the global maximum of the search.  Surprisingly, the steps on the left side of the bottom panel show that the optimizations for roughly 30\% of the grid points arrive at log-likelihood values much less than the global maximum. This further supports using eigendecomposition instead of the grid search in this example.

\subsection{Study 2: 4-state HIV model}
\label{sec:study2}

\begin{figure}
	\centering
	\includegraphics[width=1.0\linewidth]{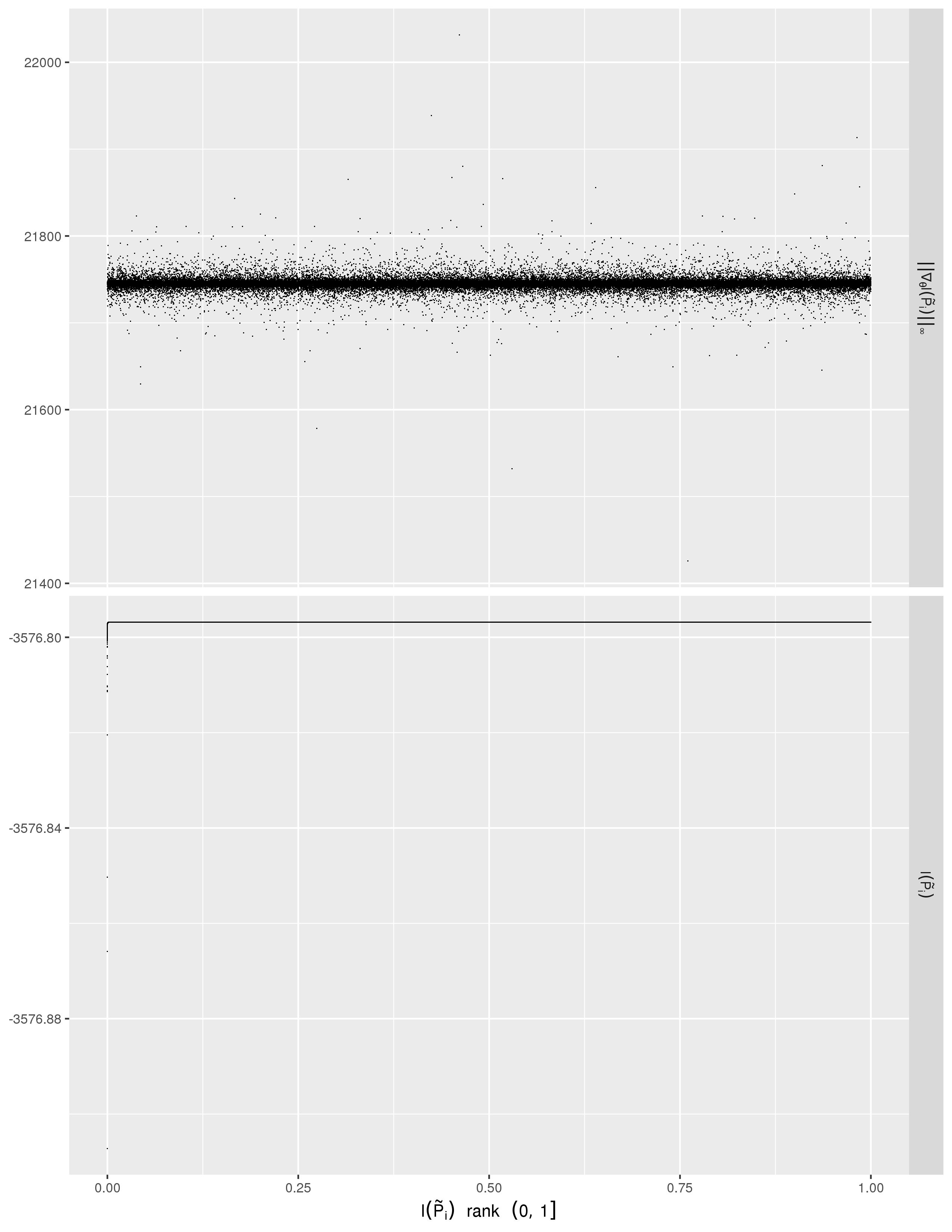}
	\caption{Bottom panel: Study 2 log-likelihood evaluated at each convergence point. Top panel: $L^\infty$ norm of gradient at each convergence point. Both panels are plotted along rank order of the log-likelihood for each convergence point (horizontal axis, $(0,1]$ scale). Any ties are ranked arbitrarily.}
	\label{fig:study2_sortedlik}
\end{figure}

The second study uses the annual HIV progression data in (\ref{eqn:chancellordata}), for which the principal 12th root of the annual MLE has a negative probability. The seven non-principal 12th roots are also not stochastic.  Optimization fails at 257,760 grid points.  Figure \ref{fig:study2_sortedlik} shows that each remaining grid point converges to approximately the same log-likelihood value with very few exceptions. The global maximizer from the search is:
\[
\widehat{\mathbf{P}} = \left[
\begin{array}{rrrr}
.973 	&	.025	&	.002	&	.000\\
0 		&	.956 	&   .044 	& 	.000\\
0		&	0		&   .978	&   .022\\
0		&	0		&	0		&		1
\end{array}
\right].
\]
The monthly MLE has effectively zero probability of transitioning from the first or second state to the fourth state.
The log-likelihood gradient at $\widehat{\mathbf{P}}$ is:
\[
\nabla_\theta l(\widehat{\mathbf{P}}) = \left[
\begin{array}{rrr}
28.421  & 69.054  & 104.774\\
0  & 21710.401 & 21745.079\\
0  &      0 &  -3837.441\\
0 & 0 & 0
\end{array}
\right].
\]  
The gradient terms are large because the free parameters in the first and second rows of $\widehat{\mathbf{P}}$ are at the constraint boundary. 

The global maximizer in this study is close to the estimate obtained by Chhatwal et al. using an approximation method \cite{chhatwal_changing_2016}.
Chhatwal et al. evaluate their approximation error as
\[
\frac{||\widehat{\mathbf{P}}^{12} - \widehat{\mathbf{P}^{12}}||_F}{||\widehat{\mathbf{P}^{12}}||_F},
\]
where $||.||_F$ is the Frobenius norm.
The error in their estimate is 3.37\%.
By the same calculation, $\widehat{\mathbf{P}}$ in this study has an error of 3.84\%.  This is roughly the same error as Chhatwal et al. with the advantage that the estimate here is an MLE.  Note, however, that this is not actually an error in the present setting because $\widehat{\mathbf{P}^{12}}$ is not a valid estimate under the assumption that a stochastic 12th root exists.  

\subsection{Study 3: 5-state HIV model}
\label{sec:study3}

Study 3 uses the six-month HIV transition data in (\ref{eqn:sendicharitosdata}). Monthly transition probabilities are not available because the six-month MLE does not have a real-valued sixth root.  Applying the method of Section \ref{sec:methods} yields the monthly MLE,
\[
\widehat{\mathbf{P}} = \left[
\begin{array}{rrrrr}
.955 & .021 & .018 & .005 & .001 \\
.201 & .670 & .106 & .023 & .000 \\
.059 & .185 & .623 & .118 & .016 \\
.027 & .000 & .078 & .861 & .034 \\
0 & 0 & 0 & 0 & 1
\end{array}
\right],
\]
with gradient, 
\[
\nabla_\theta l(\widehat{\mathbf{P}}) = \left[
\begin{array}{rrrr}
.049 &  .048 & .046 & .047\\
426.782 & 426.781 & 426.779 & 426.778\\
.013 & .012 & .010 & .010\\
.001 & -57.638 & -.001 &  .000\\
0  &   0 &    0 &      0
\end{array}
\right].
\]  
The optimization fails at 25,964 grid points.

\begin{figure}
	\centering
	\includegraphics[width=1.0\linewidth]{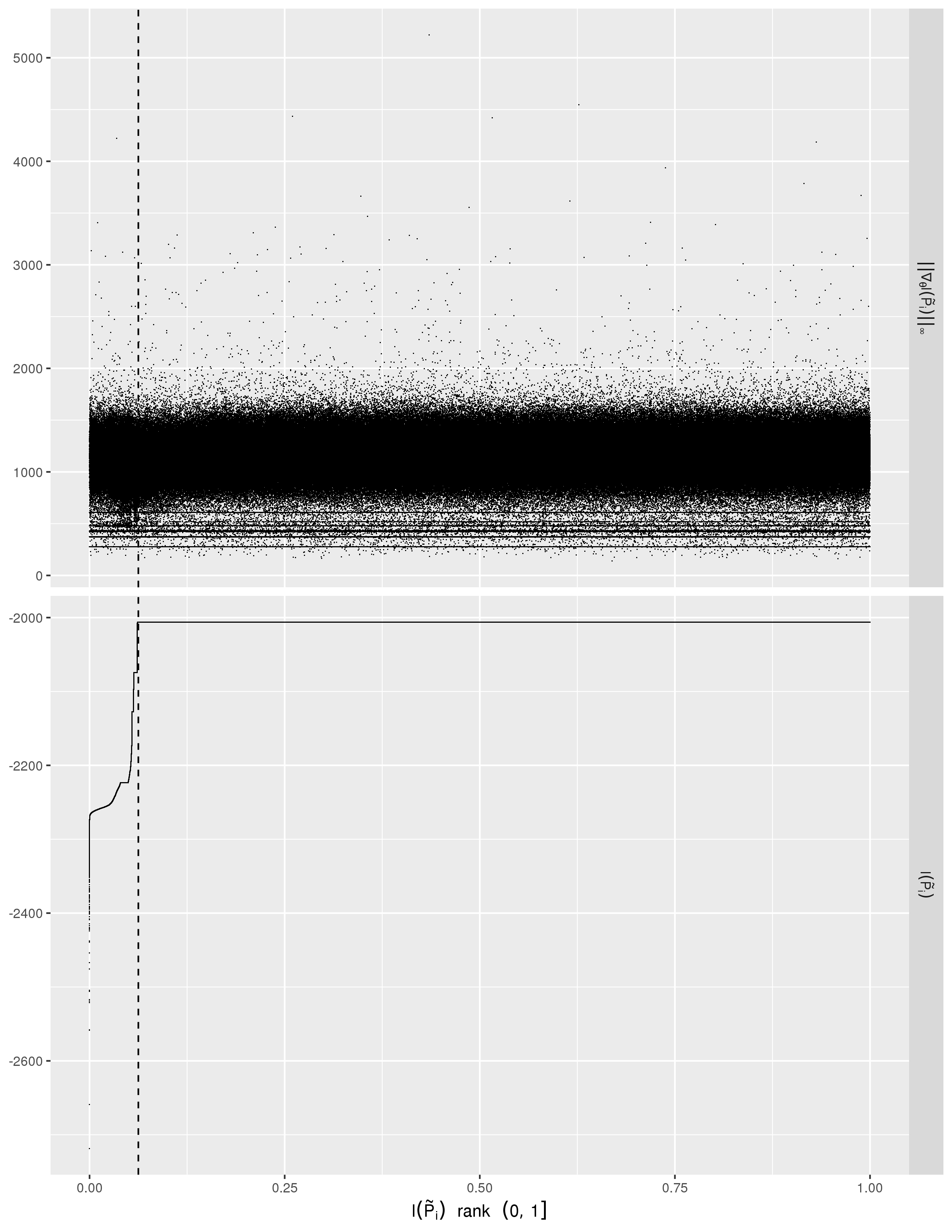}
	\caption{Bottom panel: Study 3 log-likelihood evaluated at each convergence point. Top panel: $L^\infty$ norm of gradient at each convergence point. Both panels are plotted along rank order of the log-likelihood for each convergence point (horizontal axis, $(0,1]$ scale). Any ties are ranked arbitrarily. Convergence points to the right of the dashed line are analyzed further for uniqueness.}
	\label{fig:study3_sortedlik}
\end{figure}

\begin{figure}
	\centering
	\includegraphics[width=1.0\linewidth]{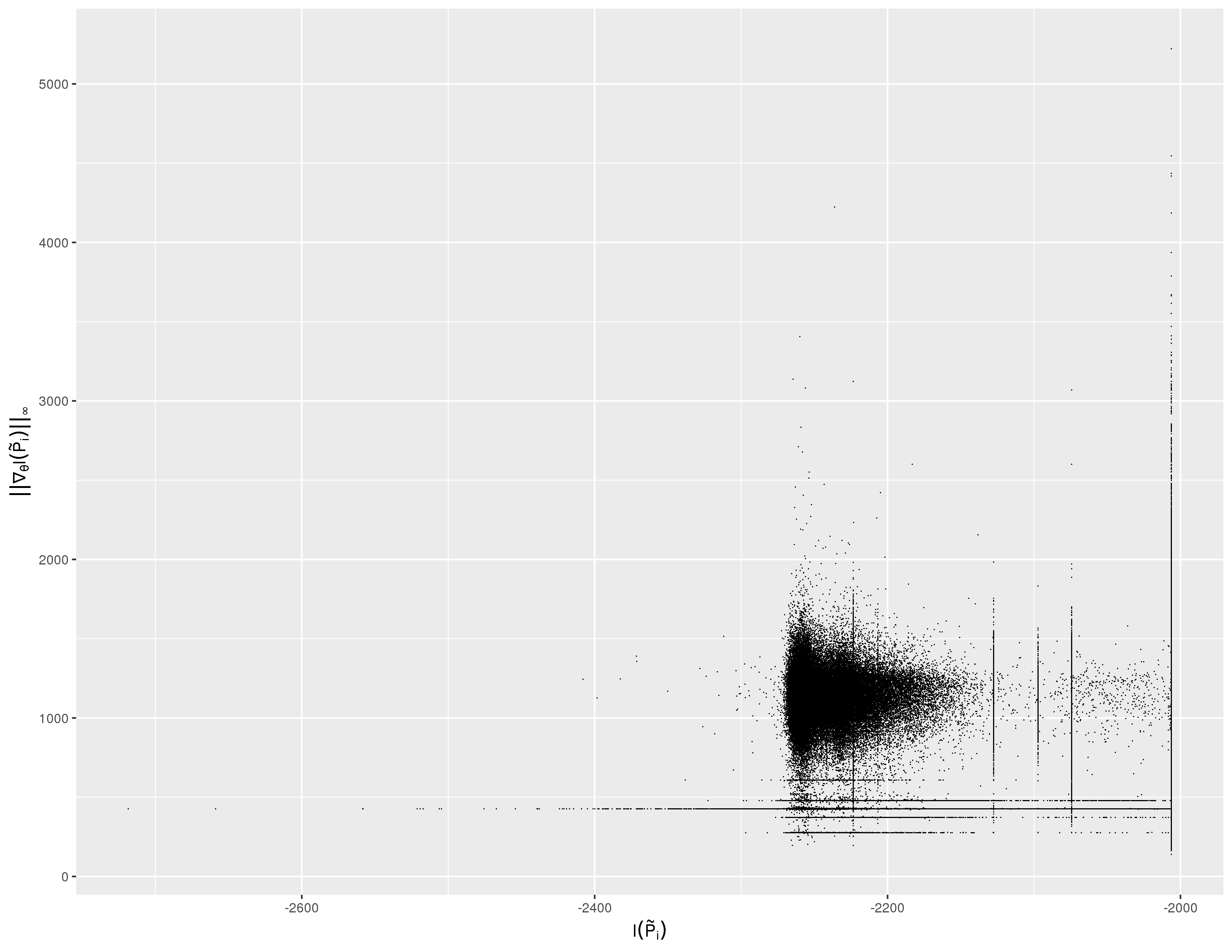}
	\caption{$L^\infty$ norm of gradient (vertical axis) and log-likelihood (horizontal axis) at each convergence point in the Study 3 grid search.}
	\label{fig:study3_gradient_likelihood}
\end{figure}

\begin{figure}
	\centering
	\includegraphics[width=1.0\linewidth]{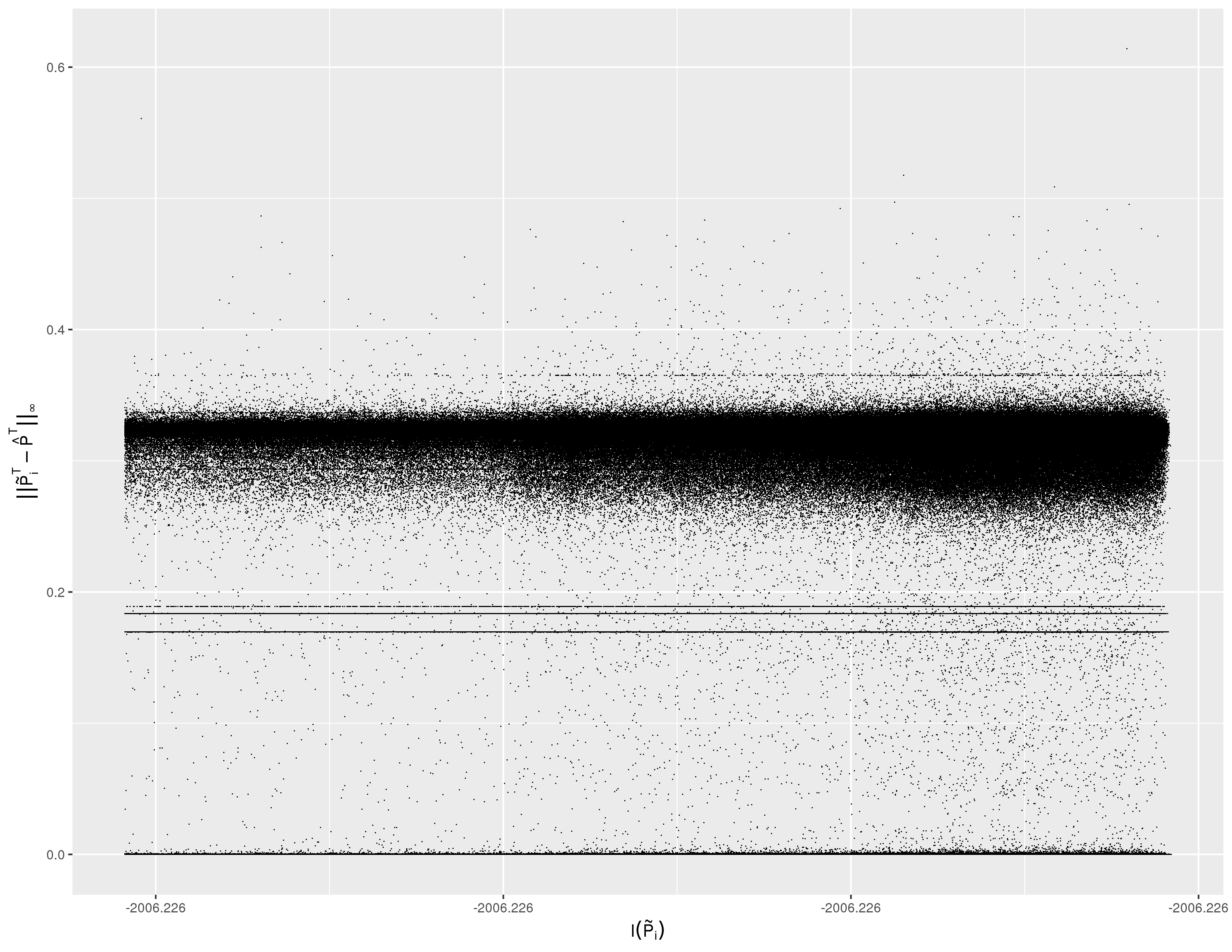}
	\caption{$L^\infty$ distances between global maximizer and convergence points in the Study 3 grid search. The horizontal axis has the corresponding log-likelihood values.  Only convergence points with the top 93.75\% of log-likelihood values are shown.}
	\label{fig:study3_linfdist}
\end{figure}

The bottom panel of Figure \ref{fig:study3_sortedlik} shows many small plateaus in which grid points converge near the same log-likelihood value.  However,  roughly 94\% of grid points converge near the global maximum of the search.  Unlike any of the other studies, the gradient never stabilizes within any convergence plateau (Figure \ref{fig:study3_sortedlik}, top panel).  The problem is illustrated more clearly in Figure \ref{fig:study3_gradient_likelihood}, where the horizontal axis is log-likelihood units instead of rank order. The regions of constant log-likelihood with apparently continuous and very disperse gradients suggest there are likelihood ridges at the boundary of the parameter space.  
Figure \ref{fig:study3_linfdist} shows the $L^\infty$ distance between the global maximizer of the search, $\widehat{\mathbf{P}}^6,$ and each convergence point, $\widetilde{\mathbf{P}}_i^6,$ among the top 93.75\% of log-likelihood values (demarcated with a dashed line in Figure \ref{fig:study3_sortedlik}).   The horizontal bands indicate multiple distinct maximizers of equal likelihood very close to the global maximum of the search.  The topmost horizontal band in Figure \ref{fig:study3_linfdist} suggests a transition probability in one of the maximizers differs by approximately 0.36 from the corresponding probability in the global maximizer. These are very different estimates with nearly equal likelihood. Note that these are multiple distinct maximizers in terms of $\mathbf{P}^6$, each of which could have multiple distinct stochastic roots.

\subsection{Study 4: 3-state chain with one negative eigenvalue}

Study 4 considers the following synthetic data:
\[
\mathbf{N} = \left[
\begin{array}{rrr}
200& 650& 400\\
350& 100& 450\\
100& 500& 250
\end{array}
\right].
\]
The MLE over the observation interval has eigenvalues $\lambda_2 = -.439,$ and $\lambda_3=.004.$ A stochastic root is unavailable for any value of $T \in \{2,24,100\}$ because $\lambda_2$ is negative.

\begin{figure}
	\centering
	\includegraphics[width=1.0\linewidth]{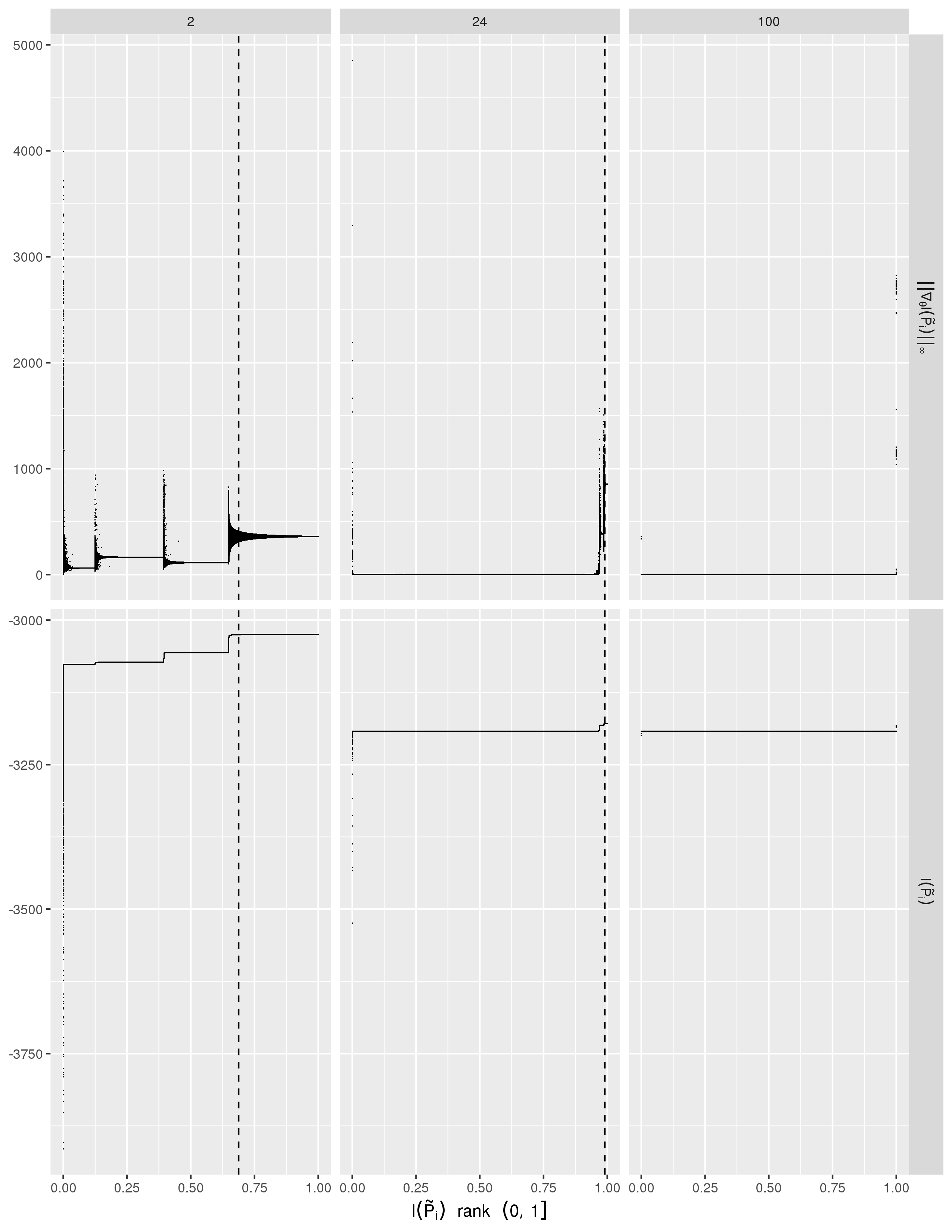}
	\caption{Bottom panels: Study 4 log-likelihood for $T=2,$ $24,$ and $100$  evaluated at each convergence point. Top panels: $L^\infty$ norm of gradient at each convergence point. All panels are plotted along rank order of the log-likelihood for each convergence point (horizontal axis, $(0,1]$ scale). Any ties are ranked arbitrarily.  Convergence points to the right of the dashed line are analyzed further for uniqueness.}
	\label{fig:study4_sortedlik}
\end{figure}

\begin{figure}
	\centering
	\includegraphics[width=1.0\linewidth]{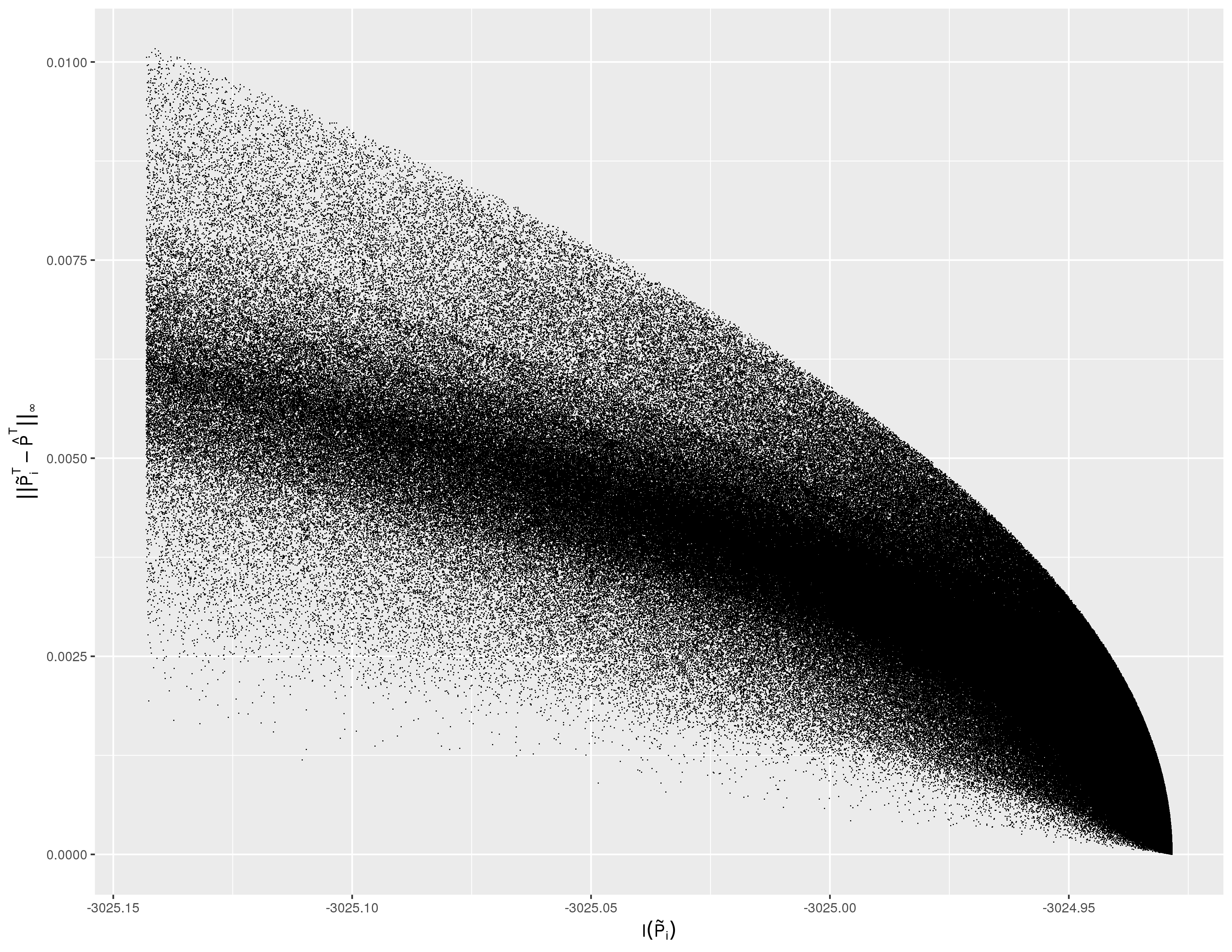}
	\caption{$L^\infty$ distances between global maximizer and convergence points in the Study 4 grid search for $T=2$. The horizontal axis has the corresponding log-likelihood values.  Only convergence points with the top 31.25\% of log-likelihood values are shown.}
	\label{fig:study4_2_linfdist}
\end{figure}

For $T=2$, the search finds several convergence plateaus after optimization fails at five grid points.  The plateaus are shown in the bottom-left panel of Figure \ref{fig:study4_sortedlik}. The gradient is nonzero in each of the plateaus (upper-left panel).  Therefore, these are convergence points at the boundary and not in flat regions in the interior.  The rightmost plateau in the bottom-left panel shows that about 30\% of the grid points converge to a likelihood near the global maximum of the search.  While the gradient stabilizes in this plateau (unlike in Study 3), the question still remains whether these points correspond to a single global maximizer in terms of $\mathbf{P}^2$ or there are multiple distinct maximizers of equal likelihood. Figure \ref{fig:study4_2_linfdist} shows the $L^\infty$ distance between the global maximizer and each convergence point among the top 31.25\% of log-likelihood values (demarcated with a dashed line in the left panels of Figure \ref{fig:study4_sortedlik}).  Elementwise, the transition probability estimates at these convergence points are all within .011 of the corresponding global maximizer estimate.
This difference goes to zero as the points approach the global maximum.
Therefore, the global maximizer for $\widehat{\mathbf{P}}^2$ appears unique.

\begin{figure}
	\centering
	\includegraphics[width=1.0\linewidth]{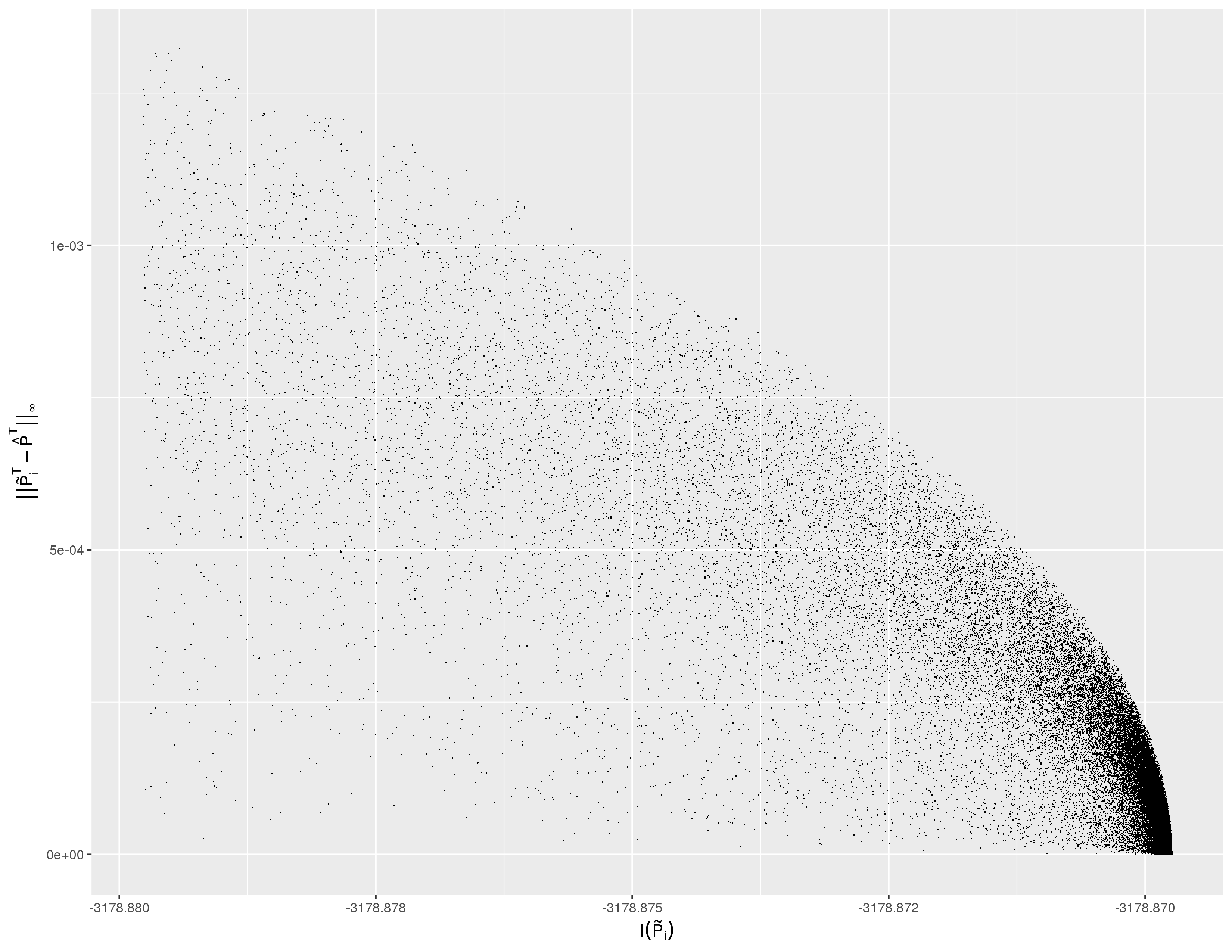}
	\caption{$L^\infty$ distances between global maximizer and convergence points in the Study 4 grid search for $T=24$. The horizontal axis has the corresponding log-likelihood values.  Only convergence points with the top 1\% of log-likelihood values are shown.}
	\label{fig:study4_24_linfdist}
\end{figure}

For $T=24$, optimization fails at two grid points and roughly 99\% of the remaining grid points reach their convergence criteria at approximately the same log-likelihood value.  This is represented by the wide plateau in the lower-middle panel of Figure \ref{fig:study4_sortedlik}.  The gradient is zero at these convergence points (upper-middle panel), suggesting the likelihood is nearly flat in the interior of the parameter space.  These convergence points are not the global maximizer of the search, however.  There are two narrow plateaus on the right side of the lower-middle panel. The highest plateau, containing the global maximizer in the search, represents about 1\% of the grid points. 
The gradient is nonzero in this plateau.  
Figure \ref{fig:study4_24_linfdist} compares the maximizers in this top 1\% to the global maximizer.  The transition probability estimates at these convergence points are all within .002 of the corresponding global maximizer estimate. The difference appears to go to zero as the convergence points' log-likelihood approaches the global maximum. Therefore, the global maximizer appears to be unique in terms of $\widehat{\mathbf{P}}^{24}$.

For $T=100$, the optimization runs to completion at all grid locations.  Nearly all locations produce convergence points with approximately the same log-likelihood value (Figure \ref{fig:study4_sortedlik}, bottom-right panel).  The gradient is zero throughout, suggesting the optimizations reach a region of nearly flat log-likelihood in the interior (Figure \ref{fig:study4_sortedlik}, top-right panel).  The two small clusters of nonzero gradient points on the rightmost side of the top-right panel indicate two convergence regions on the boundary.  However, there are so few convergence points in these clusters that the convergence plateaus near the global maximum are invisible in the bottom-right panel.

\begin{table}
	
	\centering
	
	\begin{tabular}{|c|c|c|}
		\multicolumn{1}{c}{ } & \multicolumn{1}{c}{$\widehat{\mathbf{P}}$}&
		\multicolumn{1}{c}{$\nabla_\theta \,l\left(\widehat{\mathbf{P}}\right)$} \\		
		\hline
		$T=2$ & $\left[
		\begin{array}{rrr}
		 .000 & .775 & .225\\
		 .000 & .501 & .499\\
		 .762 & .000 & .238
		\end{array}
		\right]$&
		$\left[
		\begin{array}{rr}
		-14.807 & .002\\
		-360.156 & -.008\\
	   -.002 & -6.827
		\end{array}
		\right]$\\
		\hline
		$T=24$ & $\left[
		\begin{array}{rrr}
		.858 & .000 & .142\\
		.078 & .922 & .000\\
		.000 & .089 & .911
		\end{array}
		\right]$&
		$\left[
		\begin{array}{rr}
		.009 & -227.041\\
	    22.819 &  22.790\\
		-851.854 &  -.005
		\end{array}
		\right]$		\\
		\hline
		$T=100$ & $\left[
		\begin{array}{rrr}
		.963 & .000 & .037\\
		.020 & .980 & .000\\
		.000 & .023 & .977
		\end{array}
		\right]$&
		$\left[
		\begin{array}{rr}
	    .238 & -674.966\\
         32.510 &  32.708\\
        -2736.605 & .121
		\end{array}
		\right]$		\\
		\hline
	\end{tabular}
	\caption{Global maximizers and their gradients from the Study 4 grid search.}
	\label{tab:study4_convergence}
\end{table}

\subsection{Study 5: 3-state chain with two negative eigenvalues}

\begin{figure}
	\centering
	\includegraphics[width=1.0\linewidth]{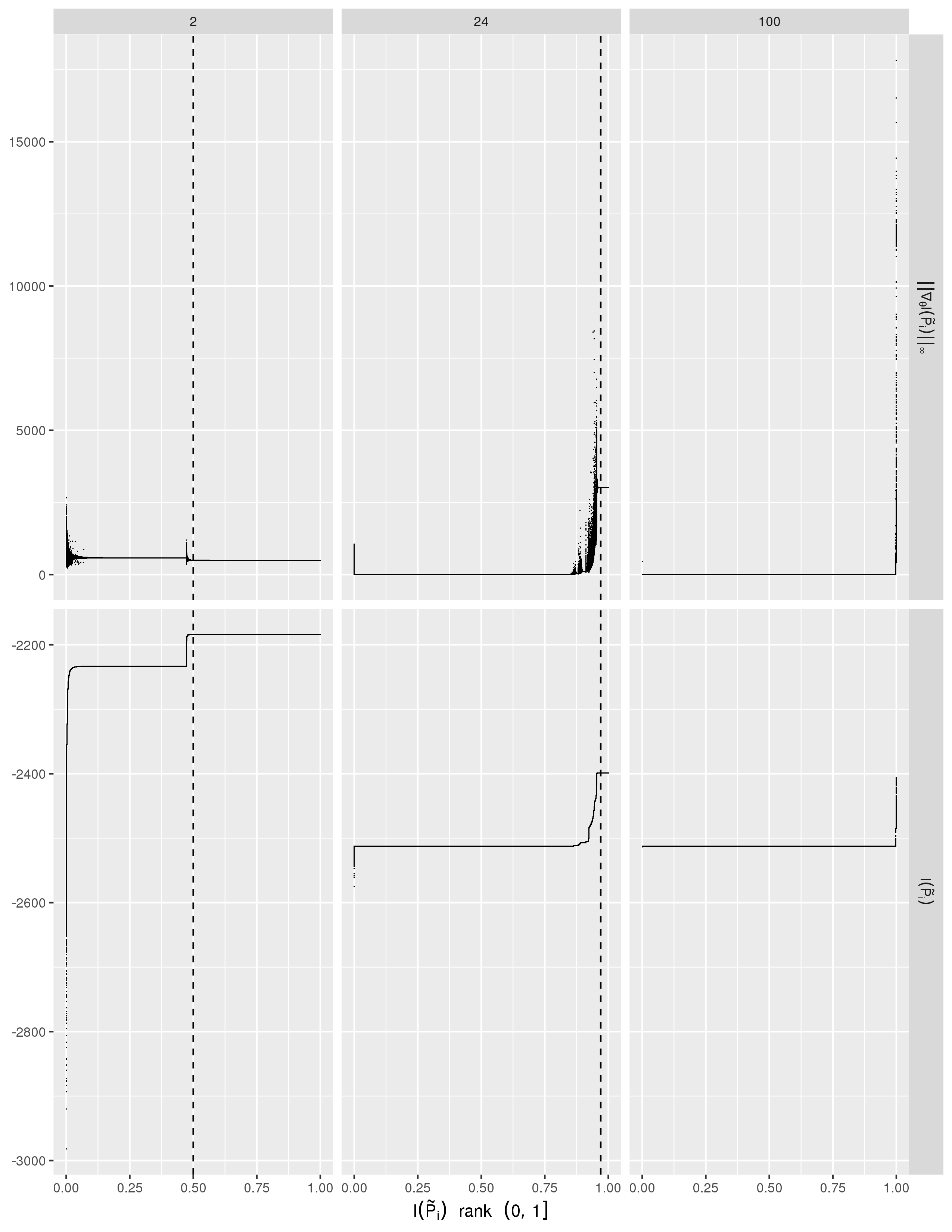}
	\caption{Bottom panels: Study 5 log-likelihood for $T=2,$ $24,$ and $100$  evaluated at each convergence point. Top panels: $L^\infty$ norm of gradient at each convergence point. All panels are plotted along rank order of the log-likelihood for each convergence point (horizontal axis, $(0,1]$ scale). Any ties are ranked arbitrarily.  Convergence points to the right of the dashed line are analyzed further for uniqueness.}
	\label{fig:study5_sortedlik}
\end{figure}

The synthetic data for the fifth study are:
\[
\mathbf{N} = \left[
\begin{array}{rrr}
100&200&650\\
300&350&100\\
250&300&50
\end{array}
\right].
\]
The observation interval MLE has no stochastic roots due to two negative eigenvalues ($\lambda_2 = -.328$ and $\lambda_3=-.017$).  Table \ref{tab:study5_convergence} shows the global maxima of the search and their gradients.  Optimization fails at 13, 56, and 30 grid points for $T=2,$ $24,$ and $100,$ respectively.  The bottom panels of Figure \ref{fig:study5_sortedlik} show the log-likelihood values where each grid point's optimization converges.  Like in Study 4, there are fewer grid points in the convergence plateau around the global maximizer as $T$ increases and, for $T=100$, there is no visible plateau.  Also, for $T=24$ and $T=100$, most grid points appear to meet their convergence criteria in an interior region.  For $T=2$ and $T=24$, around 50\% and 3\% of points in the respective grids converge to a log-likelihood near the global maximum.  The probabilities in the top 50\% for $T=2$ are all within .003 of the global maximum estimate.  The top 3\% are within .001 for $T=24$.  The differences within each plateau are shown in Supporting Figures \ref{fig:study5_2_linfdist} and \ref{fig:study5_24_linfdist}, respectively. These small, decreasing differences indicate unique global maximizers for $T=2$ and $T=24$ in terms of $\mathbf{P}^T$. 

\begin{table}
	
	\centering
	
	\begin{tabular}{|c|c|c|}
		\multicolumn{1}{c}{ } & \multicolumn{1}{c}{$\widehat{\mathbf{P}}$}&
		\multicolumn{1}{c}{$\nabla_\theta \,l\left(\widehat{\mathbf{P}}\right)$} \\		
		\hline
		$T=2$ & $\left[
		\begin{array}{rrr}
		.206 & .794 & .000\\
		.181 & .043 & .775\\
		.591 & .248 & .161
		\end{array}
		\right]$&
		$\left[
		\begin{array}{rr}
		491.270 & 491.270\\
		.000 & -.001\\
	   .001  &  .001
		\end{array}
		\right]$\\
		\hline
		$T=24$ & $\left[
		\begin{array}{rrr}
		.919 & .000 & .081\\
		.039 & .961 & .000\\
		.018 & .049 & .933
		\end{array}
		\right]$&
		$\left[
		\begin{array}{rr}
	  -.001 & -2011.865\\
	   3014.974 &  3014.990\\
	  -.011   &    -.002
		\end{array}
		\right]$		\\
		\hline
		$T=100$ & $\left[
		\begin{array}{rrr}
		.980 & .000 & .020\\
		.010 & .990 & .000\\
	    .005 & .012 & .983
		\end{array}
		\right]$&
		$\left[
		\begin{array}{rr}
	    .004  & -7743.587\\
	    11652.493 & 11652.492\\
	    -.017 &   -.002
		\end{array}
		\right]$		\\
		\hline
	\end{tabular}
	\caption{Global maximizers and their gradients from the Study 5 grid search.}
	\label{tab:study5_convergence}
\end{table} 

\subsection{Study 6: 3-state chain with complex eigenvalues}

The sixth study involves a complex conjugate eigenvalue pair.  The synthetic data are:
\[
\mathbf{N} = \left[
\begin{array}{rrr}
200&400&100\\
100&250&300\\
150&200&100
\end{array}
\right].
\]
The observation interval MLE is,
\[
\widehat{\mathbf{P}^T} = \left[
\begin{array}{rrr}
.286 & .571 & .143\\
.154 & .385 & .462\\
.333 & .444 & .222
\end{array}
\right],
\]
with eigenvalues  ($\lambda_2 = -.054+.151\mathrm{i}$ and $\lambda_3 = -.054-.151\mathrm{i}$). 
There are two real-valued roots when $T=2$,
\[
\left[
\begin{array}{rrr}
.601 & .560  & -.160\\
-.032 & .502 & .530\\
.357 & .284 &  .359
\end{array}
\right] \mbox{ and } 
\left[
\begin{array}{rrr}
-.118 & .337 & .781\\
.515 &  .395 & .090\\
.126 & .612 & .262
\end{array}
\right],
\]
neither of which are stochastic.
None of the roots for $T=24$ and $T=100$ are stochastic either.

\begin{figure}
	\centering
	\includegraphics[width=1.0\linewidth]{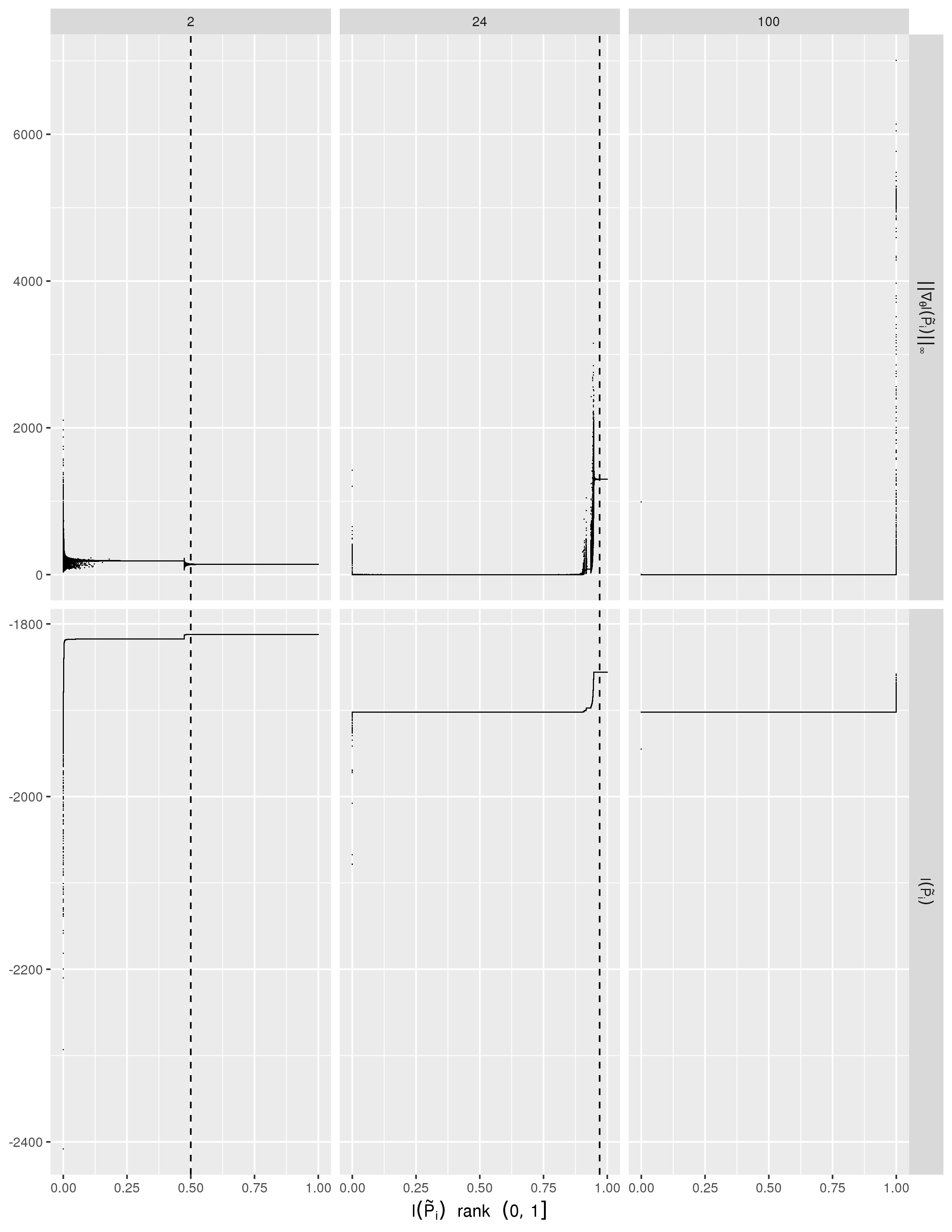}
	\caption{Bottom panels: Study 6 log-likelihood for $T=2,$ $24,$ and $100$  evaluated at each convergence point. Top panels: $L^\infty$ norm of gradient at each convergence point. All panels are plotted along rank order of the log-likelihood for each convergence point (horizontal axis, $(0,1]$ scale). Any ties are ranked arbitrarily.  Convergence points to the right of the dashed line are analyzed further for uniqueness.}
	\label{fig:study6_sortedlik}
\end{figure}

The global maximizers of the search and the gradients are shown in Table \ref{tab:study6_convergence}.  Optimization fails at 4, 27, and 12 grid points for $T=2,$ $24,$ and $100,$ respectively.
The convergence results in this study (Figure \ref{fig:study6_sortedlik}) are very similar to Study 5 (Figure \ref{fig:study5_sortedlik}).  About 50\% of the grid points converge near the global maximum for $T=2$ and about 3\% for $T=24$.  There is no visible convergence plateau near the maximum for $T=100.$  
The probabilities in the top 50\% of convergence points for $T=2$ are within .003 of the global maximum estimate. For $T=24,$ the top 3\% are within .001 of the global maximizer.  Supporting Figures \ref{fig:study6_2_linfdist} and \ref{fig:study6_24_linfdist} show the respective differences within each of these plateaus. In both cases the differences vanish approaching the global maximum.
Thus, the convergence plateaus for $T=2$ and $T=24$ appear to represent unique maximizers.

\begin{table}
	
	\centering
	
	\begin{tabular}{|c|c|c|}
		\multicolumn{1}{c}{ } & \multicolumn{1}{c}{$\widehat{\mathbf{P}}$}&
		\multicolumn{1}{c}{$\nabla_\theta \,l\left(\widehat{\mathbf{P}}\right)$} \\		
		\hline
		$T=2$ & $\left[
		\begin{array}{rrr}
		.000 & .317 & .683\\
		.505 & .345 & .150\\
		.160 & .684 & .156
		\end{array}
		\right]$&
		$\left[
		\begin{array}{rr}
		-140.262 &    .000\\
	    .000 &    .000\\
	    .000 &   .000
		\end{array}
		\right]$\\
		\hline
		$T=24$ & $\left[
		\begin{array}{rrr}
		.938 & .062 & .000 \\
		.000 & .954 & .046 \\
		.051 & .021 & .927
		\end{array}
		\right]$&
		$\left[
		\begin{array}{rr}
		1299.309 & 1299.308\\
		-827.211 &   .001\\
		-.001  & -.003
		\end{array}
		\right]$		\\
		\hline
		$T=100$ & $\left[
		\begin{array}{rrr}
		.985 & .015 & .000\\
		.000 & .989 & .011\\
		.013 & .005 & .982
		\end{array}
		\right]$&
		$\left[
		\begin{array}{rr}
		5076.655 & 5076.649\\
		-3310.540  &  .015\\
	    .005 &   .004 \\
		\end{array}
		\right]$		\\
		\hline
	\end{tabular}
	\caption{Global maximizers and their gradients from the Study 6 grid search.}
	\label{tab:study6_convergence}
\end{table}

\subsection{Study 7: 4-state chain with negative eigenvalue}

Studies 7 and 8 explore four-state chains.  Study 7 has no roots because of a negative eigenvalue.  The synthetic data are:
\[
\mathbf{N} = \left[
\begin{array}{rrrr}
100 & 200 & 650 & 100\\
300 & 350 & 100 & 200\\
250 & 300 &  50 & 300\\
100 & 200 & 300 & 400
\end{array}
\right].
\]
The observation interval MLE has eigenvalues: $\lambda_2 = -.310,$ $\lambda_3 = .222,$ and $\lambda_4 = .007.$

\begin{figure}
	\centering
	\includegraphics[width=1.0\linewidth]{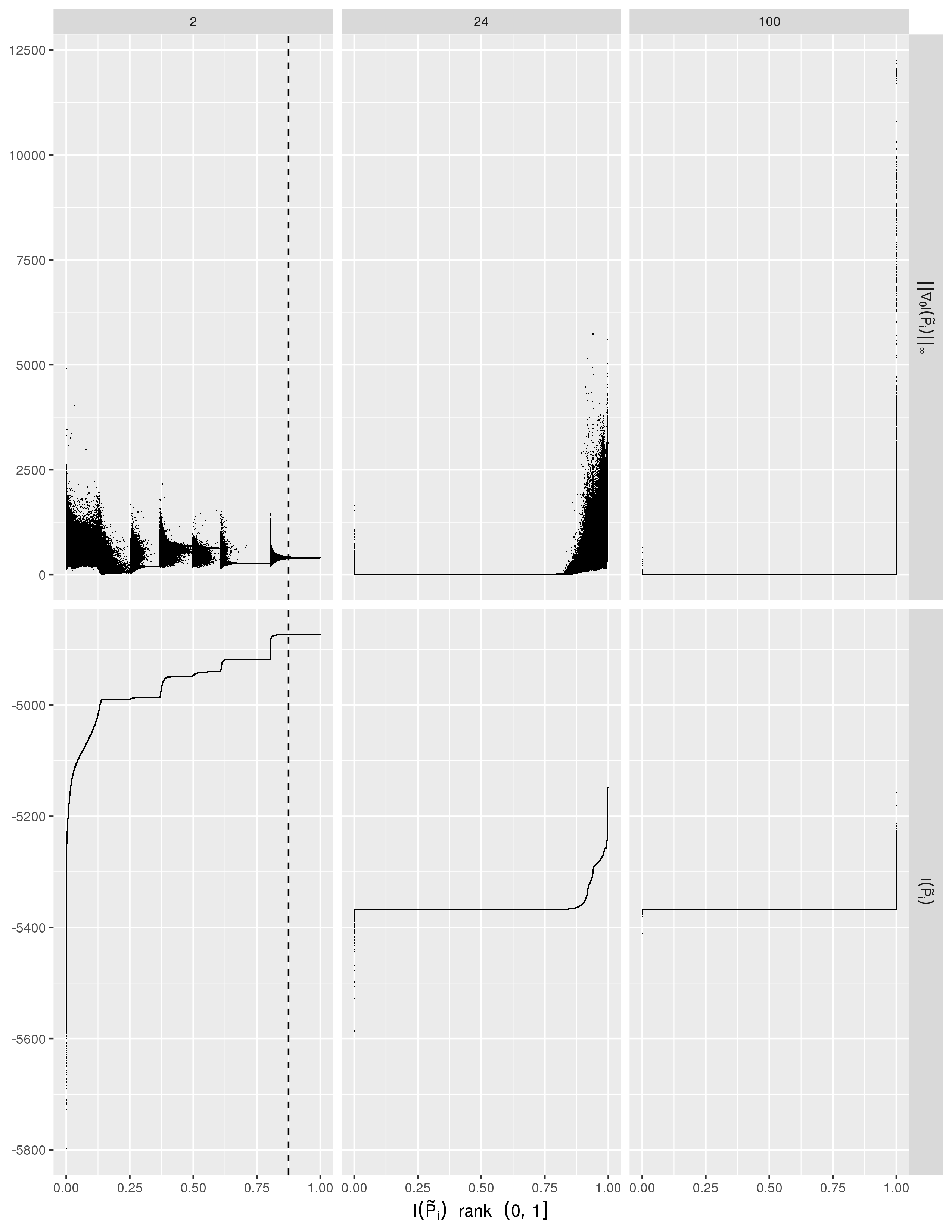}
	\caption{Bottom panels: Study 7 log-likelihood for $T=2,$ $24,$ and $100$  evaluated at each convergence point. Top panels: $L^\infty$ norm of gradient at each convergence point. All panels are plotted along rank order of the log-likelihood for each convergence point (horizontal axis, $(0,1]$ scale). Any ties are ranked arbitrarily.  Convergence points to the right of the dashed line are analyzed further for uniqueness.}
	\label{fig:study7_sortedlik}
\end{figure}

Table \ref{tab:study7_convergence} shows the global maximizers and gradients for this study.  Optimization fails at 11 grid points for $T=2$ and $24$, and at one grid point for $T=100.$
There appear to be at least six convergence plateaus for $T=2$ (Figure \ref{fig:study7_sortedlik}, bottom-left panel), many more than in the synthetic $s=3$ studies (Studies 4--6). The $T=24$ convergence plateaus (Figure \ref{fig:study7_sortedlik}, lower-middle panel) are so closely spaced that they appear to be a continuous curve. For $T=24$ and $T=100$, nearly all grid points converge to a zero-gradient region where the log-likelihood is less than the global maximum.  For $T=2,$ about 13\% of the grid points converge near the global maximum of the search.  Transition probabilities among this top 13\% are within .005 of their global MLE (Supporting Figure \ref{fig:study7_2_linfdist}).  Given the small differences that are decreasing on approach to the maximum likelihood, the global maximizer of the $T=2$ search seems to be unique.   

\begin{table}
	
	\centering
	\scalebox{0.94}{	
	\begin{tabular}{|c|c|c|}
		\multicolumn{1}{c}{ } & \multicolumn{1}{c}{$\widehat{\mathbf{P}}$}&
		\multicolumn{1}{c}{$\nabla_\theta \,l\left(\widehat{\mathbf{P}}\right)$} \\		
		\hline
		$T=2$ & $\left[
		\begin{array}{rrrr}
		.210 & .715 & .000 & .075\\
		.175 & .000 & .773 & .053\\
		.407 & .146 & .154 & .294\\
		.046 & .338 & .000 & .616
		\end{array}
		\right]$&
		$\left[
		\begin{array}{rrr}
		.036 &   .026 & -402.074\\
		.021 & -132.976 &   .008\\
		.036 &    .036 &   .025\\
		-.005 &  -.007 &  -44.290
		\end{array}
		\right]$\\
		\hline
		$T=24$ & $\left[
		\begin{array}{rrrr}
		.919 & .000 & .081 & .000\\
		.034 & .949 & .000 & .017\\
		.009 & .046 & .919 & .026\\
		.011 & .009 & .026 & .955
		\end{array}
		\right]$&
		$\left[
		\begin{array}{rrr}
		1870.668 & -31.145 &  1870.723\\
		.018 & -.096 & -3129.985\\
	   -.065 & -.098 &    .010\\
	   -.088 & -.073 &    .070
		\end{array}
		\right]$		\\
		\hline
		$T=100$ & $\left[
		\begin{array}{rrrr}
		 .980 & .000 & .020 & .000\\
		 .008 & .987 & .000 & .004\\
		 .002 & .011 & .980 & .006\\
		 .003 & .002 & .007 & .989
		\end{array}
		\right]$&
		$\left[
		\begin{array}{rrr}
		7244.156 & -14.832  & 7244.209\\
		-0.019  & -0.062 & -11980.129\\
		 0.031 & -0.008  &    0.059\\
	   -0.072  & -0.109  &   -0.012
		\end{array}
		\right]$		\\
		\hline
	\end{tabular}
	}
	\caption{Global maximizers and their gradients from the Study 7 grid search.}
	\label{tab:study7_convergence}
\end{table}

\subsection{Study 8: 4-state chain with negative and complex eigenvalues}

The synthetic data for Study 8 are:
\[
\mathbf{N} = \left[
\begin{array}{rrrr}
200 & 650 & 400 & 100\\
350 & 100 & 450 & 200\\
100 & 500 & 250 & 300\\
400 & 300 & 200 & 100
\end{array}
\right].
\]
The observation interval MLE has both negative and complex eigenvalues ($\lambda_2 = -.357,$ $\lambda_3 = -.043+.170\mathrm{i},$ and $\lambda_4 = -.043-.170\mathrm{i}$).  

\begin{figure}
	\centering
	\includegraphics[width=1.0\linewidth]{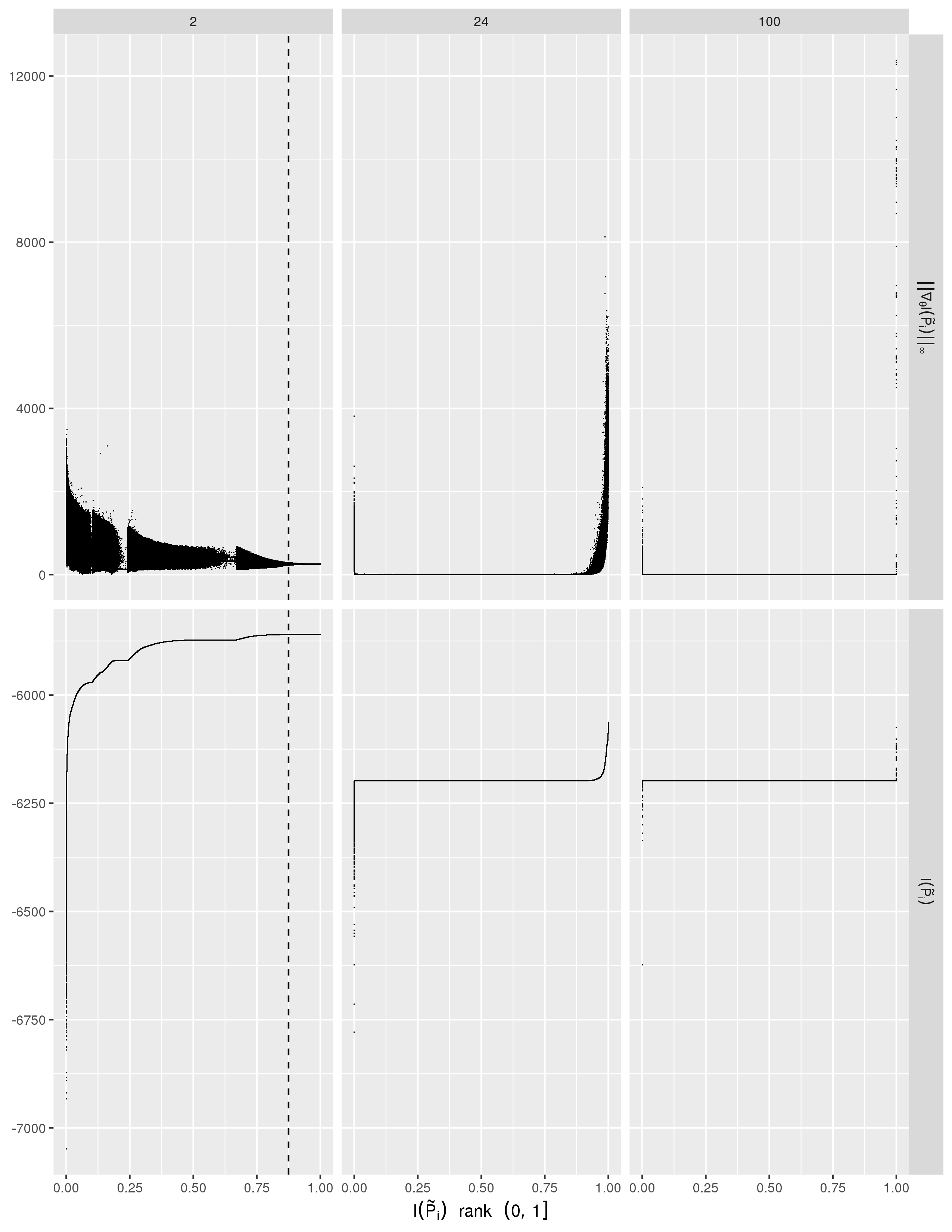}
	\caption{Bottom panels: Study 8 log-likelihood for $T=2,$ $24,$ and $100$  evaluated at each convergence point. Top panels: $L^\infty$ norm of gradient at each convergence point. All panels are plotted along rank order of the log-likelihood for each convergence point (horizontal axis, $(0,1]$ scale). Any ties are ranked arbitrarily.  Convergence points to the right of the dashed line are analyzed further for uniqueness.}
	\label{fig:study8_sortedlik}
\end{figure}

Table \ref{tab:study8_convergence} shows the global maximizers and gradients.  
Optimization fails at 3 and 7 grid points for $T=2$ and $24,$  respectively. None of the optimizations fail for $T=100.$
Unlike the other studies, the plateaus in the lower-left panel of Figure \ref{fig:study8_sortedlik} ($T=2$) are so closely spaced that the majority of the convergence points appear to be sitting on a curve. Neither the $T=24$ nor $T=100$ searches find a plateau around the global maximum.  For $T=2$, around 13\% of the points converge near the global maximum. That global maximum plateau for $T=2$ is more difficult to find than in previous studies because its log-likelihood value is not much greater than the next largest plateau.  The transition probabilities within this top 13\% are within .012 of their global maximizer estimates (Supporting Figure \ref{fig:study8_2_linfdist}). The differences go to zero approaching global maximum.  The plateau for $T=2$ seems to correspond to a unique maximizer.

\begin{table}
	
	\centering
	\scalebox{0.90}{	
	\begin{tabular}{|c|c|c|}
		\multicolumn{1}{c}{ } & \multicolumn{1}{c}{$\widehat{\mathbf{P}}$}&
		\multicolumn{1}{c}{$\nabla_\theta \,l\left(\widehat{\mathbf{P}}\right)$} \\		
		\hline
		$T=2$ & $\left[
				\begin{array}{rrrr}
				 .025 & .301 & .211 & .463\\
				 .215 & .369 & .417 & .000\\
				 .540 & .000 & .153 & .307\\
				 .000 & .694 & .306 & .000
				\end{array}
				\right]$&
				$\left[
				\begin{array}{rrr}
				 -.069 &  -.007 & -.041\\
				 253.566 & 253.544 & 253.558\\
				 .047 & -131.636 &  .007\\
				 -73.725 &  130.756  & 130.749
				\end{array}
				\right]$\\
		\hline
		$T=24$ & $\left[
		\begin{array}{rrrr}
		.919 & .000 & .081 & .000\\
		.000 & .906 & .039 & .055\\
		.000 & .108 & .892 & .000\\
		.107 & .000 & .000 & .893
		\end{array}
		\right]$&
		$\left[
		\begin{array}{rrr}
		2645.061 & 2380.088 & 2645.052\\
		-562.497 &  -.054  & -.080\\
		1381.186 & 3325.479 & 3325.464\\
		-.012 & -718.366 & -626.450
		\end{array}
		\right]$		\\
		\hline
		$T=100$ & $\left[
		\begin{array}{rrrr}
		 .980 & .000 & .020 & .000\\
		 .000 & .976 & .010 & .014\\
		 .000 & .028 & .972 & .000\\
		 .027 & .000 & .000 & .973
		 \end{array}
		\right]$&
		$\left[
		\begin{array}{rrr}
		10188.296 &  9246.027 & 10188.303\\
		-2233.645 &   -.002   & -.001\\
		4948.569  & 12326.711 & 12326.716\\
		-.005 & -2790.863 & -2648.348
		\end{array}
		\right]$		\\
		\hline
	\end{tabular}
	}
	\caption{Global maximizers and their gradients from the Study 8 grid search.}
	\label{tab:study8_convergence}
\end{table}

\section{Discussion}
\label{sec:discussion}

Enumeration of real-valued roots (Section \ref{sec:rootfinding}) is only feasible for moderately sized $s$ and $T$ like in Studies 6 and 8.  When larger $s$ and $T$ make this computationally impractical, the root finding step should be skipped. The stochastic roots are maxima to be discovered in the grid search.  

From the results of Studies 4--8, it appears that the number of convergence plateaus increases with $s.$ 
For $T=2$ the convergence plateaus around the global maximum of the searches are narrower for $s=4$ than for $s=3$.  Also, the $s=3$ studies (4--6) have  convergence plateaus near the global maximizer for $T=24.$  Plateaus are absent for $T=24$ in the $s=4$ studies (7 and 8).  Altogether, the studies suggest the global maximum is more difficult to locate as $s$ increases.

Within each study, the global maximizer is more difficult to locate as $T$ increases.  For $T=100$ in Studies 4--8 and $T=24$ in Studies 7 and 8, at most a handful of grid points converge near the global maximum of the search.  These would be more convincing global maximizers if a visible plateau of grid points converged near that log-likelihood value.
For this reason, when there is no visible convergence plateau near the global maximum of an initial search, an iteratively refined search should be performed within some radius of the global maximizer until a convincing plateau is realized or a greater maximum is discovered.

The simplicity of direct maximization suggested by MacDonald \cite{macdonald2014} seems to come mainly from 
not implementing the gradient explicitly.  In \R version 3.6.3, this requires either (a) coding each element of $\mathbf{P}^T$ as a function of the elements in $\mathbf{P}$ and using an algorithmic derivative or (b) using a numerical difference approximation. For large $T$, coding the individual $\mathbf{P}^T$ terms has no obvious advantage over the method presented here.  It is not clear how numerical derivatives could fail in practice.  There may be practical examples where numerical approximation errors lead to poorer performance than analytic approaches.  This is a question for future research.

Craig and Sendi's EM algorithm requires enumerating all possible state transition paths for the cycles in the observation interval and then computing the expected number of transitions in each E-step \cite{craig_estimation_2002}. Their M-step optimization is simple compared to the present method because their objective is concave and the gradient does not require computing $\partial \mathbf{P}^T/\partial \theta_{uv}$ in (\ref{eqn:delPT_delp}).  Therefore, when analytical gradients are used, it is not clear if the direct optimization presented here is computationally more efficient than their EM approach.  The question boils down to the convergence rates of the two algorithms and the amount of computation in their E-step versus the gradient evaluation in (\ref{eqn:loglikgradient}).  This, too, remains to be studied.

The present work assumes $T$ is an integer.  An extension to rational $T$ is straightforward \cite{higham_pth_2011}. If $T=q/r$, treat $\mathbf{P}^{1/r}$ as the parameter for likelihood maximization with observation interval length $q$ in order to obtain $\widehat{\mathbf{P}^{1/r}}$.  Then, the cycle-length MLE is $(\widehat{\mathbf{P}^{1/r}})^r.$  However, this is potentially undesirable because it assumes the Markov structure holds for cycles as short as $1/r$, whereas, for integer $T$, the Markov structure is assumed only for unit-length cycles.

The approximation methods in Table \ref{tab:methods_classification} work directly with transition matrices whereas likelihood-based methods require transition counts.  Clearly, the approximation methods are useful when transition counts are not available.  
However, those who publish a transition matrix with no stochastic root  are implicitly assuming that the Markov structure is irrelevant for a given cycle length.  Analysts are then forced to work around this assumption by approximation.  It would be more valuable if the data provider created an analysis process whereby analysts have indirect access to transition counts or work in a secure data enclave with controlled access.  This would allow richer and perhaps more realistic model development.

Ultimately, the grid search requires a great deal of computational effort and, for $T=24$ and $T=100$, the initial effort does not always produce a convincing global maximizer.  
This effort is shaped by four requirements: not assuming embeddability, time-homogeneity, data observed at a fixed time interval, and likelihood maximization.
Are these important enough to justify the effort?  
What alternatives become available by relaxing each requirement?

When the embeddability assumption is justified, then \cite{kalbfleisch_analysis_1985} has maximum likelihood estimation methods for both time-homogeneous and inhomogeneous chains. However, it is not clear whether the methods in \cite{kalbfleisch_analysis_1985} have any computational advantages over the present approach.

In many contexts, time-homogeneity is useful only as a local approximation.  For example, chronic disease risk typically increases with age.  A constant risk assumption might be defensible for monthly cycles within a single year, but would be implausible over, say, a decade.
Time-homogeneity is also dubious in the credit rating context \cite{israel_finding_2001,kreinin_regularization_2001}.
Regression models allow time dependencies into the single-cycle transition probabilities \cite{erminileaf_2017unified}.

The present model assumes observations occur at discrete time points while ignoring the exact time of transitions. However, continuous timing is observable with relative ease in many contexts.  Survey participants can be asked the date of a recent disease diagnosis or it can be obtained from medical records.  Credit ratings change as relevant events occur \cite{spglobalguide}.  
Survival models estimated from continuous time data can be used to compute transition probabilities for any cycle length.
For example, Chancellor et al.  compute annual transition probabilities from a survival model \cite{chancellor_modelling_1997}.  If available, that survival model could be used to compute the monthly transition probabilities directly instead of approximating the $12$th root of the annual matrix as in \cite{chhatwal_changing_2016}.

Finally, the benefits of an asymptotic distribution for the MLE are not readily available when the MLE is on the parameter boundary as in Studies 2--8 \cite{self_asymptotic_1987}.  Additional work is needed to derive this.  For inference purposes, Bayesian approaches (see Table \ref{tab:methods_classification}) or a constrained inferential model \cite[Chapter~5]{martin2015inferential} could be mathematically and computationally simpler alternatives to likelihood maximization.

\section*{Acknowledgements}
The author acknowledges the Center for Advanced Research Computing (CARC) at the University of Southern California for providing computing resources that have contributed to the research results reported within this publication. URL: https://carc.usc.edu.

\bibliography{direct_MLE_of_transition_matrix}

\begin{thebibliography}{10}

\bibitem{jahn_alternative_2019}
Jahn B, Kurzthaler C, Chhatwal J, Elbasha EH, Conrads-Frank A, Rochau U, et~al.
\newblock Alternative conversion methods for transition probabilities in
  state-transition models: validity and impact on comparative effectiveness and
  cost-effectiveness.
\newblock Med Decis Making. 2019;39(5):509--522.
\newblock \url{https://doi.org/10.1177/0272989X19851095}.

\bibitem{chhatwal_changing_2016}
Chhatwal J, Jayasuriya S, Elbasha EH.
\newblock Changing cycle lengths in state-transition models: challenges and
  solutions.
\newblock Med Decis Making. 2016;36(8):952--964.
\newblock \url{https://doi.org/10.1177/0272989X16656165}.

\bibitem{chancellor_modelling_1997}
Chancellor JV, Hill AM, Sabin CA, Simpson KN, Youle M.
\newblock Modelling the cost effectiveness of {Lamivudine}/{Zidovudine}
  combination therapy in {HIV} infection.
\newblock Pharmacoeconomics. 1997 Jul;12(1):54--66.
\newblock \url{https://doi.org/10.2165/00019053-199712010-00006}.

\bibitem{sendi_estimating_1999}
Sendi PP, Bucher HC, Craig BA, Pfluger D, Battegay M.
\newblock Estimating {AIDS}-free survival in a severely immunosuppressed
  asymptomatic {HIV}-infected population in the era of antiretroviral triple
  combination therapy.
\newblock J Acquir Immune Defic Syndr Hum Retrovirol. 1999;20(4):376--381.
\newblock \url{https://doi.org/10.1097/00042560-199904010-00008}.

\bibitem{charitos_computing_2008}
Charitos T, de~Waal PR, van~der Gaag LC.
\newblock Computing short-interval transition matrices of a discrete-time
  {Markov} chain from partially observed data.
\newblock Stat Med. 2008;27(6):905--921.
\newblock \url{https://doi.org/10.1002/sim.2970}.

\bibitem{kingman_imbedding_1962}
Kingman JFC.
\newblock The imbedding problem for finite {Markov} chains.
\newblock Zeitschrift für Wahrscheinlichkeitstheorie und verwandte Gebiete.
  1962 Jan;1:14--24.
\newblock \url{https://doi.org/10.1007/BF00531768}.

\bibitem{lin_roots_2011}
Lin L.
\newblock Roots of stochastic matrices and fractional matrix powers.
\newblock Manchester, UK: University of Manchester; 2011.
\newblock \url{https://www.proquest.com/docview/1787508874}.

\bibitem{israel_finding_2001}
Israel RB, Rosenthal JS, Wei JZ.
\newblock Finding generators for {Markov} chains via empirical transition
  matrices, with applications to credit ratings.
\newblock Math Financ. 2001;11(2):245--265.
\newblock \url{https://doi.org/10.1111/1467-9965.00114}.

\bibitem{briggs_model_2012}
Briggs AH, Weinstein MC, Fenwick EAL, Karnon J, Sculpher MJ, Paltiel AD.
\newblock Model parameter estimation and uncertainty: a report of the
  {ISPOR}-{SMDM} {Modeling} {Good} {Research} {Practices} {Task} {Force}-6.
\newblock Value Health. 2012 Sep;15(6):835--842.
\newblock \url{https://doi.org/10.1016/j.jval.2012.04.014}.

\bibitem{craig_estimation_2002}
Craig BA, Sendi PP.
\newblock Estimation of the transition matrix of a discrete-time {Markov}
  chain.
\newblock Health Econ. 2002;11(1):33--42.
\newblock \url{https://doi.org/10.1002/hec.654}.

\bibitem{macdonald2014}
MacDonald IL.
\newblock Numerical maximisation of likelihood: a neglected alternative to
  {EM}?
\newblock Int Stat Rev. 2014;82(2):296--308.
\newblock \url{https://doi.org/10.1111/insr.12041}.

\bibitem{bladt_statistical_2005}
Bladt M, S{\o}rensen M.
\newblock Statistical inference for discretely observed {Markov} jump
  processes.
\newblock J R Stat Soc Series B Stat Methodol. 2005;67(3):395--410.
\newblock \url{https://doi.org/10.1111/j.1467-9868.2005.00508.x}.

\bibitem{kalbfleisch_analysis_1985}
Kalbfleisch JD, Lawless JF.
\newblock The analysis of panel data under a {Markov} assumption.
\newblock J Am Stat Assoc. 1985;80(392):863--871.
\newblock \url{https://doi.org/10.1080/01621459.1985.10478195}.

\bibitem{welton_estimation_2005}
Welton NJ, Ades AE.
\newblock Estimation of {Markov} chain transition probabilities and rates from
  fully and partially observed data: uncertainty propagation, evidence
  synthesis, and model calibration.
\newblock Med Decis Making. 2005;25(6):633--645.
\newblock \url{https://doi.org/10.1177/0272989X05282637}.

\bibitem{kreinin_regularization_2001}
Kreinin A, Sidelnikova M.
\newblock Regularization algorithms for transition matrices.
\newblock Algo Research Quarterly. 2001 Jun;4(1/2):23--40.
\newblock \url{https://www.researchgate.net/profile/Alexander-Kreinin/publication/215992315_Regularization_Algorithms_for_Transition_Matrices/links/0912f50d38c9e9ca02000000/Regularization-Algorithms-for-Transition-Matrices.pdf}.

\bibitem{coleman_introduction_1964}
Coleman JS.
\newblock Introduction to Mathematical Sociology.
\newblock New York: Free Press of Glencoe; 1964.

\bibitem{crommelin_fitting_2006}
Crommelin DT, Vanden-Eijnden E.
\newblock Fitting timeseries by continuous-time {Markov} chains: a quadratic
  programming approach.
\newblock J Comput Phys. 2006;217(2):782--805.
\newblock \url{https://doi.org/10.1016/j.jcp.2006.01.045}.

\bibitem{singer_representation_1976}
Singer B, Spilerman S.
\newblock The representation of social processes by {Markov} models.
\newblock AJS. 1976;82(1):1--54.
\newblock \url{https://doi.org/10.1086/226269}.

\bibitem{zahl_markov_1955}
Zahl S.
\newblock A {Markov} process model for follow-up studies.
\newblock Hum Biol. 1955 May;27(2):90--120.
\newblock \url{https://www.proquest.com/docview/1301826510}.

\bibitem{higham_pth_2011}
Higham NJ, Lin L.
\newblock On {\it p}th roots of stochastic matrices.
\newblock Linear Algebra Appl. 2011 Aug;435(3):448--463.
\newblock \url{https://doi.org/10.1016/j.laa.2010.04.007}.

\bibitem{lange2010}
Lange K.
\newblock Numerical analysis for statisticians.
\newblock 2nd ed. New York: Springer; 2010.
\newblock \url{https://link.springer.com/content/pdf/10.1007/978-1-4419-5945-4.pdf}.

\bibitem{Rlang}
{R Core Team}. R: a language and environment for statistical computing.
\newblock Vienna, Austria: R Foundation for Statistical Computing; 2020.
\newblock \url{https://www.R-project.org/}.

\bibitem{boyd_vandenberghe_2004}
Boyd S, Vandenberghe L.
\newblock Convex optimization.
\newblock New York: Cambridge University Press; 2004.
\newblock \url{https://web.stanford.edu/~boyd/cvxbook/bv_cvxbook.pdf}.

\bibitem{data.table}
Dowle M, Srinivasan A. data.table: extension of `data.frame'; 2021.
\newblock R package version 1.14.0.
\newblock \url{https://CRAN.R-project.org/package=data.table}.

\bibitem{erminileaf_2017unified}
Ermini~Leaf D. Unified method for {Markov} chain transition model estimation
  using incomplete survey data.
\newblock ArXiv; 2017.
\newblock \url{https://arxiv.org/abs/1707.02548}.

\bibitem{spglobalguide}
{Standard \& Poor's Financial Services LLC}. Guide to credit rating essentials:
  what are credit ratings and how do they work?; 2019.
\newblock \url{https://www.spglobal.com/ratings/\_division-assets/pdfs/guide\_to\_credit\_rating\_essentials\_digital.pdf}.

\bibitem{self_asymptotic_1987}
Self SG, Liang KY.
\newblock Asymptotic properties of maximum likelihood estimators and likelihood
  ratio tests under nonstandard conditions.
\newblock J Am Stat Assoc. 1987 Jun;82(398):605--610.
\newblock \url{https://doi.org/10.1080/01621459.1987.10478472}.

\bibitem{martin2015inferential}
Martin R, Liu C.
\newblock Inferential models: reasoning with uncertainty.
\newblock New York: Chapman and Hall/CRC; 2015.
\newblock \url{https://doi.org/10.1201/b19269}.

\end{thebibliography}

\newpage
\setcounter{page}{1}
\appendix
\renewcommand{\thefigure}{S.\arabic{figure}}
\setcounter{figure}{0}

\section*{Supporting figures for \emph{\fulltitle}}

\begin{figure}[h]
	\centering
	\includegraphics[width=1.0\linewidth]{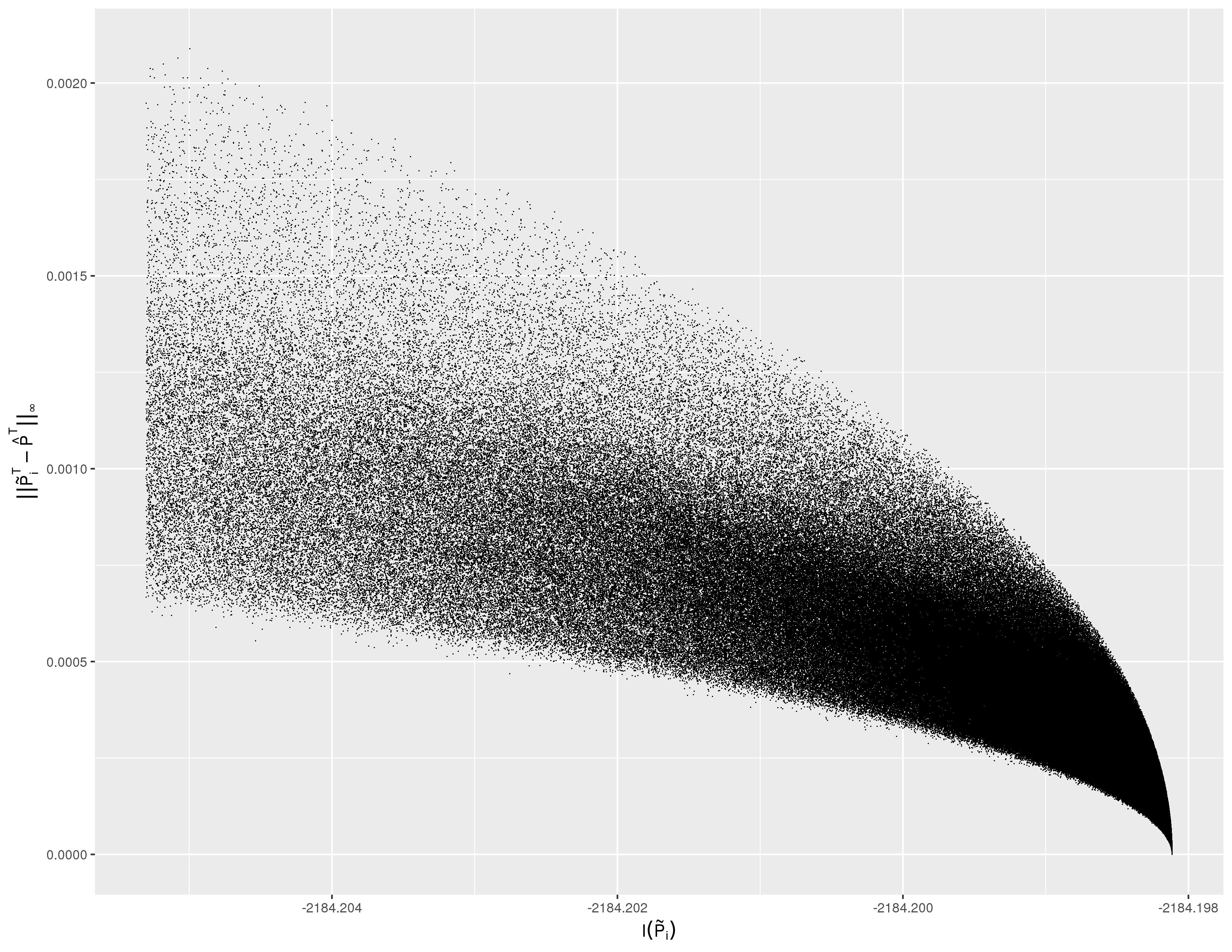}
	\caption{$L^\infty$ distances between global maximizer and convergence points in the Study 5 grid search for $T=2$. The horizontal axis has the corresponding log-likelihood values.  Only convergence points with the top 50\% of log-likelihood values are shown.}
	\label{fig:study5_2_linfdist}
\end{figure}

\begin{figure}[h]
	\centering
	\includegraphics[width=1.0\linewidth]{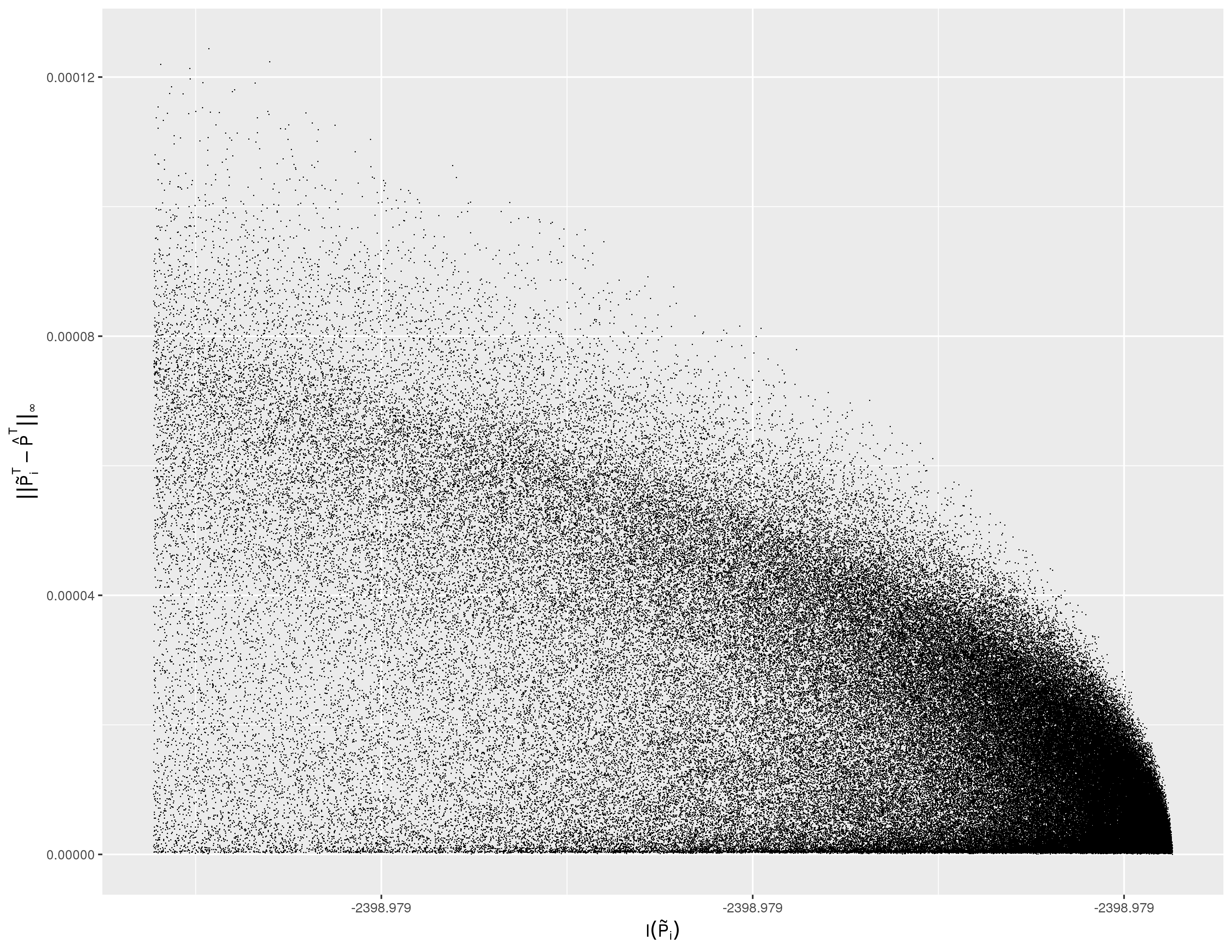}
	\caption{$L^\infty$ distances between global maximizer and convergence points in the Study 5 grid search for $T=24$. The horizontal axis has the corresponding log-likelihood values.  Only convergence points with the top 3\% of log-likelihood values are shown.}
	\label{fig:study5_24_linfdist}
\end{figure}

\begin{figure}[h]
	\centering
	\includegraphics[width=1.0\linewidth]{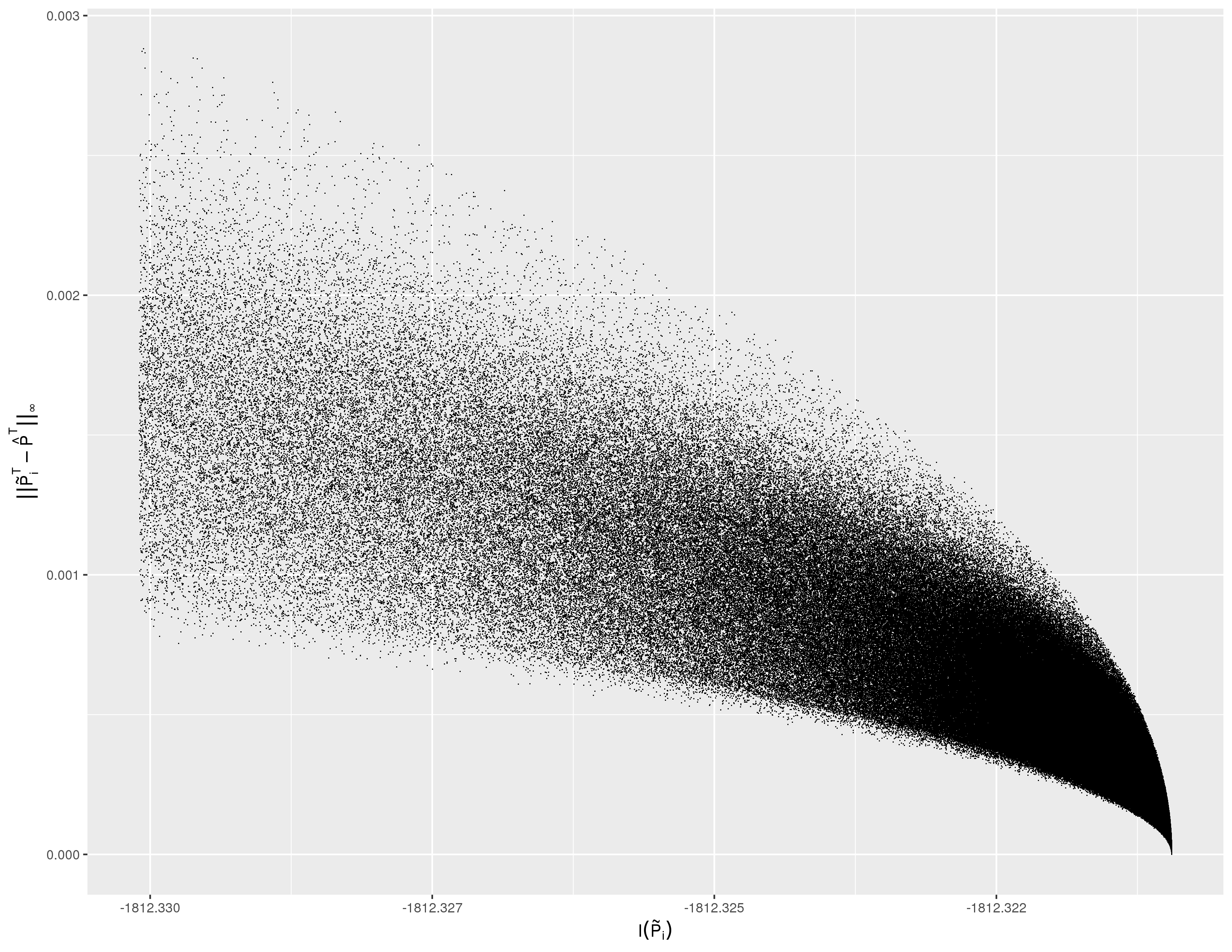}
	\caption{$L^\infty$ distances between global maximizer and convergence points in the Study 6 grid search for $T=2$. The horizontal axis has the corresponding log-likelihood values.  Only convergence points with the top 50\% of log-likelihood values are shown.}
	\label{fig:study6_2_linfdist}
\end{figure}

\begin{figure}[h]
	\centering
	\includegraphics[width=1.0\linewidth]{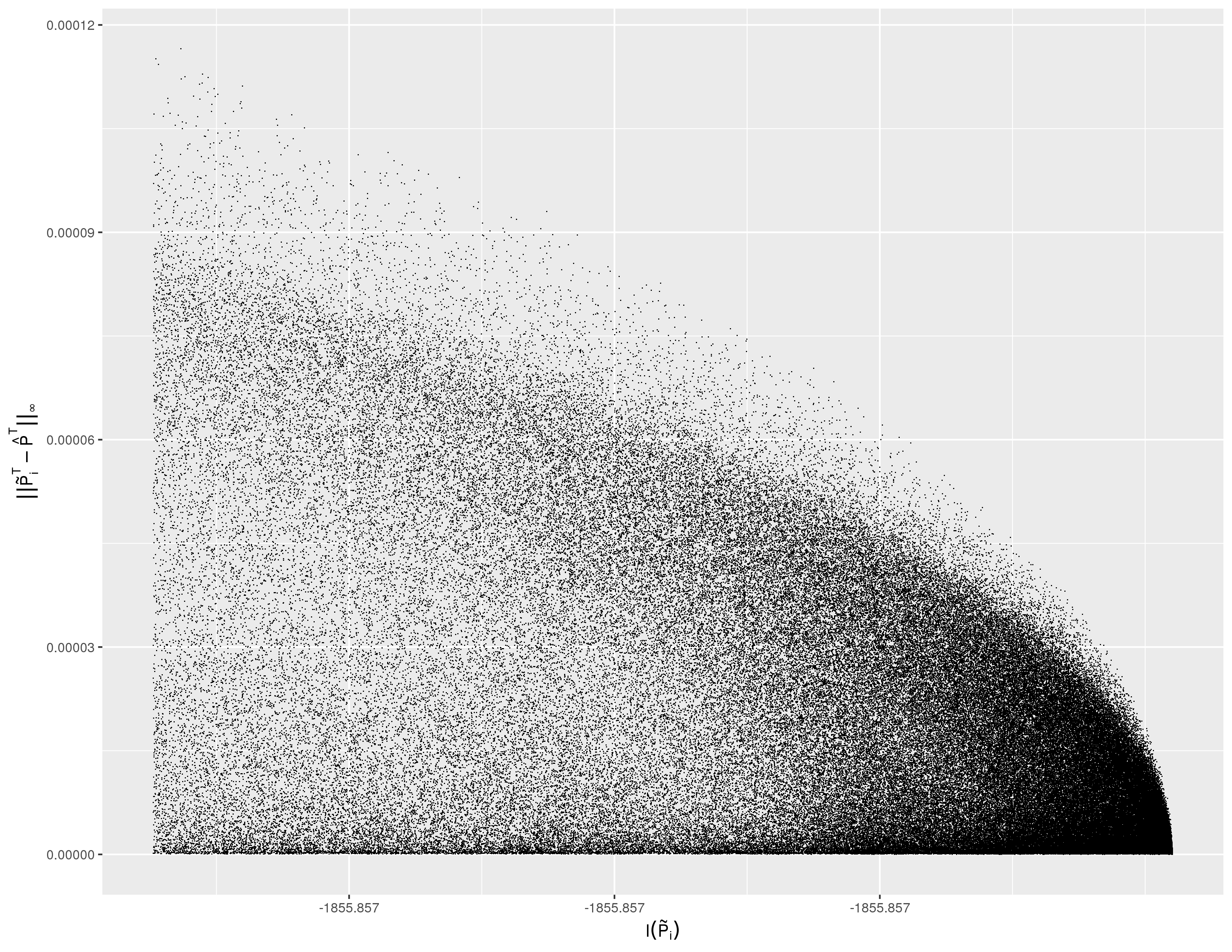}
	\caption{$L^\infty$ distances between global maximizer and convergence points in the Study 6 grid search for $T=24$. The horizontal axis has the corresponding log-likelihood values.  Only convergence points with the top 3\% of log-likelihood values are shown.}
	\label{fig:study6_24_linfdist}
\end{figure}

\begin{figure}[h]
	\centering
	\includegraphics[width=1.0\linewidth]{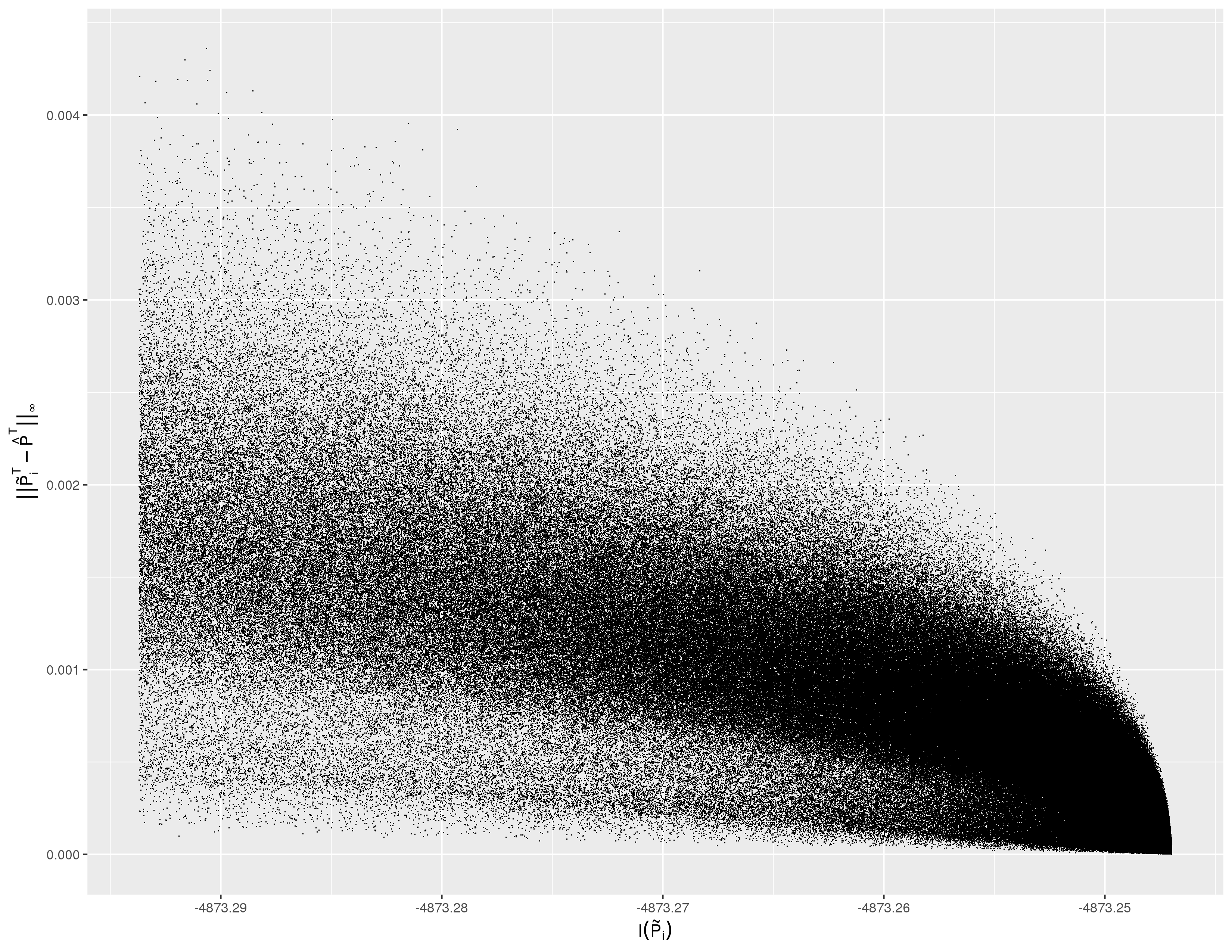}
	\caption{$L^\infty$ distances between global maximizer and convergence points in the Study 7 grid search for $T=2$. The horizontal axis has the corresponding log-likelihood values.  Only convergence points with the top 12.5\% of log-likelihood values are shown.}
	\label{fig:study7_2_linfdist}
\end{figure}

\begin{figure}[h]
	\centering
	\includegraphics[width=1.0\linewidth]{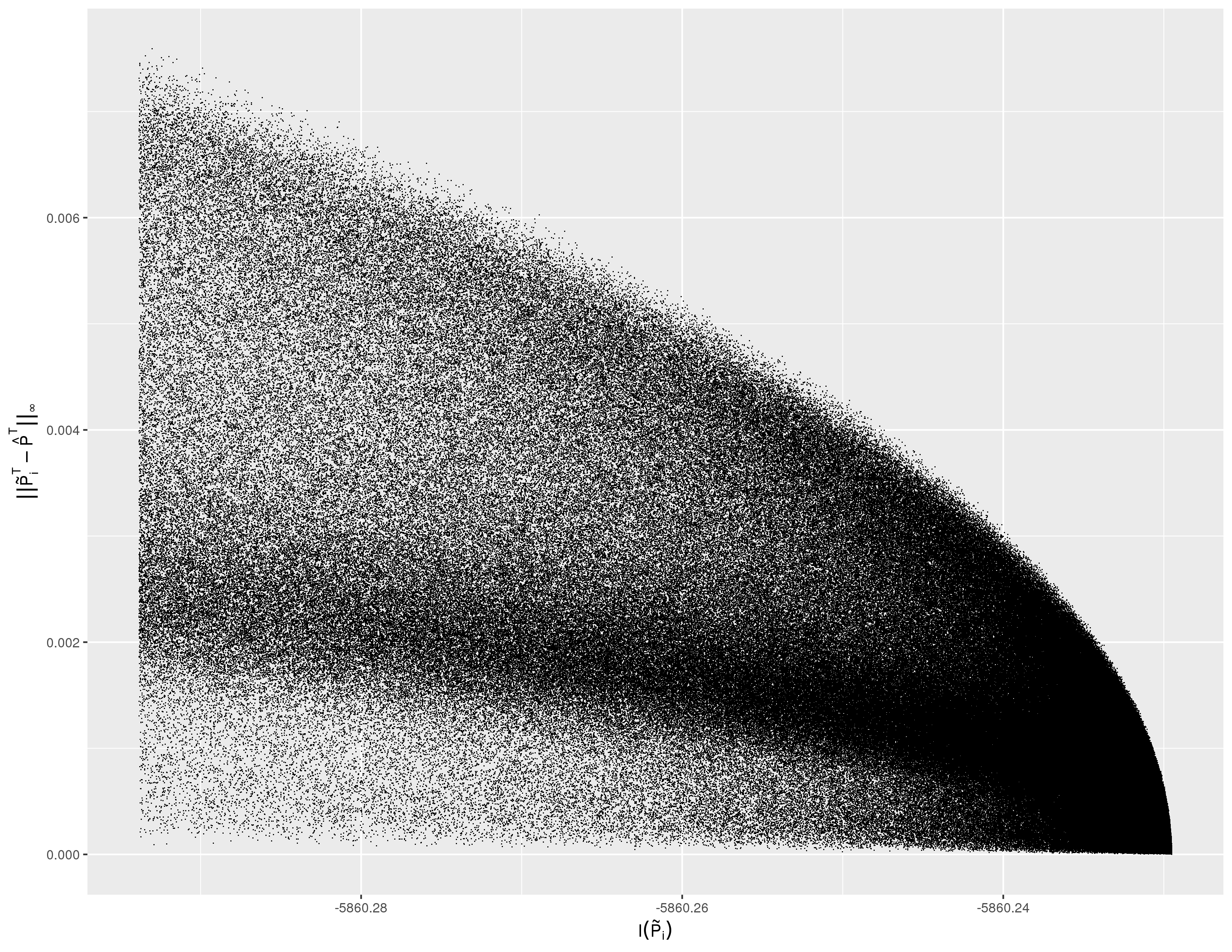}
	\caption{$L^\infty$ distances between global maximizer and convergence points in the Study 8 grid search for $T=2$. The horizontal axis has the corresponding log-likelihood values.  Only convergence points with the top 12.5\% of log-likelihood values are shown.}
	\label{fig:study8_2_linfdist}
\end{figure}

\end{document}